\documentclass[letterpaper]{mn2e}

\usepackage{rotating}
\usepackage{psfig}

\def\lsim{~\rlap{$<$}{\lower 1.0ex\hbox{$\sim$}}}
\def\bsim{~\rlap{$>$}{\lower 1.0ex\hbox{$\sim$}}}

\def\hmpc{\ {\rm {\it h}^{-1}Mpc}}

\def\mdh{\ {\rm M_\odot/{\it h}}}

\def\la{\langle}
\def\ra{\rangle}
\def\dd{{\rm d}}
\def\ln{{\rm ln}}
\def\tr{{\rm tr}}
\def\det{{\rm det}}

\def\mathbi#1{\textbf{\em #1}}

\def\rb{\bar{\rho}_m}
\def\zf{z_{\rm form}}
\def\mzf{\bar{z}_{\rm form}}
\def\ve{\mathbi{$\xi$}}

\def\nvh{\hat{\mathbi{n}}}

\def\kh{\hat{k}}

\def\rh{\hat{r}}
\def\vk{\mathbi{k}}
\def\vq{\mathbi{q}}
\def\vr{\mathbi{r}}
\def\vx{\mathbi{x}}

\def\vvs{{\bf S}}
\def\vaa{{\rm A}}
\def\vbb{{\rm B}}
\def\vcc{{\rm C}}
\def\vii{{\rm I}}
\def\vmm{{\rm M}}
\def\vqq{{\rm Q}}
\def\vrr{{\rm R}}

\def\vxx{{\rm X}}
\def\grad{\mathbi{$\nabla$}}
\def\epsi{\epsilon}
\def\dsc{\delta_{\rm sc}}
\def\dec{\delta_{\rm ec}}

\def\hw{\hat{W}}
\def\dkk{\Delta_\delta^2(k)}

\def\etal{{\it et al.\ }}

\begin{document}

\title[Environmental dependence in the ellipsoidal collapse model] 
      {Environmental dependence in the ellipsoidal collapse model}

\author[Vincent Desjacques]{Vincent Desjacques \\ 
Racah Institute of Physics, The Hebrew University, Jerusalem 91904, 
Israel \\ E-mail:~dvince@phys.huji.ac.il}

\date{}
\pagerange{\pageref{firstpage}--\pageref{lastpage}}
\maketitle

\label{firstpage}

\begin{abstract}

N-body simulations have demonstrated a correlation between the
properties of haloes and their environment. In this paper, we assess
whether the ellipsoidal collapse model, whose dynamics includes the
tidal shear,  can produce a similar dependence. First, we explore
the statistical correlation that originates from Gaussian  initial
conditions.  We derive analytic expressions for a  number of joint
statistics  of the shear tensor and estimate the sensitivity of the
local characteristics of the shear to the global geometry of the
large scale environment.  Next, we concentrate on the dynamical aspect
of the  environmental dependence using a  simplified model that takes
into account the interaction between a  collapsing halo and its
environment. We find  that the tidal force exerted by the  surrounding
mass distribution alters the axes collapse and  causes haloes embedded
in overdense regions to virialize earlier.  The environment density is
the key parameter in determining  the virialization redshift, while
the environment asphericity primarily contributes to the increase in
the  scatter of the critical collapse density.  An effective density
threshold whose shape depends on  the large scale density provides a
good description of this environmental effect. Such an interpretation
has the advantage that the excursion set formalism can be applied to
quantify the environmental dependence of halo properties. We show
that, using this approach, a correlation  between formation redshift,
large scale bias and environment  density naturally arises.  The
strength of the effect is comparable, albeit smaller, to that seen in
simulations.  It is largest for low mass haloes ($M\ll M_\star$), and
decreases as one goes to higher mass objects ($M>M_\star$).
Furthermore, haloes that formed early are substantially more clustered
than those that assembled recently.  On the other hand, our analytic
model predicts a decrease in median formation redshift  with
increasing environment density, in disagreement with the trend
detected in overdense regions.  However,  our results appear
consistent with the behaviour  inferred in relatively  underdense
regions.  We argue that the ellipsoidal collapse model may apply in
low density  environments where nonlinear effects are negligible.

\end{abstract}


\section{Introduction}
\label{sec:intro}

In standard scenarios of structure formation, dark matter haloes grow
hierarchically from initially small, Gaussian fluctuations.
Properties of haloes can be studied in great detail using both  N-body
simulations and analytic models. The remarkably useful Extended
Press-Schechter (EPS) theory predicts halo mass functions (Press \&
Schechter 1974; Bond \etal 1991), merging histories (Lacey \& Cole
1993; Sheth \& Lemson 1999b; Van den Bosch 2002; Neistein, Van den
Bosch \& Dekel 2006) and spatial clustering (Mo \& White 1996; Mo,
Jing \& White 1997; Catelan \etal 1998; Sheth \& Lemson 1999a) that
are in reasonable agreement with the simulations. This analytic
approach is based on the spherical collapse model (Gunn \& Gott 1972).
In this Lagrangian approximation, haloes are identified in the initial
conditions and a single parameter, the initial density contrast, is
needed to characterise their epoch of formation (Press \& Schechter
1974). The collapse of a halo occurs when the linear density reaches a
critical  threshold. The fundamental properties of dark matter haloes
are then obtained from the statistics of trajectories of the linear
density field as a function of the smoothing scale (e.g.  Bower 1991;
Bond \etal 1991; Lacey \& Cole 1993; Kauffmann \& White  1993;
Kitayama \& Suto 1996; Sheth \& van de Weygaert 2004).  
Although this spherical approximation works
well until the first orbit crossing, it may  not be accurate since
perturbations in Gaussian density fields are inherently triaxial
(Doroshkevich 1970; Bardeen \etal 1986; Jing \& Suto
2002). Furthermore, the initial shear field  rather than the density
has been shown to play a crucial role in the  formation of nonlinear
structures (e.g. Hoffman 1986, 1988; Peebles 1990; Dubinski 1992;
Bertschinger \& Jain 1994; Audit \&  Alimi 1996; Audit, Teyssier \&
Alimi 1997).

Unlike the spherical model whose dynamics depends on a single
parameter only (the density), the ellipsoidal model that follows the
evolution  of triaxial perturbations can be used to ascertain the
influence  of the (external) tidal shear on the properties of
collapsed regions. The gravitational collapse  of homogeneous
ellipsoids has been investigated by numerous authors over the past
decades (e.g. Lynden-Bell 1964; Lin, Mestel \& Shu 1965; Fujimoto
1968; Zeldovich 1970; Icke 1973; White \& Silk 1979; Barrow \& Silk
1981; Lemson 1993; Eisenstein \& Loeb 1995; Hui \& Bertschinger
1996). In the formulation of Bond \& Myers (1996), initial conditions
and external tides are chosen to recover the Zeldovich approximation
in the linear regime. The dynamic of ellipsoidal collapse can be
incorporated in various ways in the Press-Schechter  formalism to
predict the properties of haloes  (Monaco 1995, 1997a, 1997b; Lee \&
Shandarin 1998; Chiueh \& Lee 2001; Sheth \& Tormen 2002). As pointed
out by Sheth, Mo \& Tormen (2001), the inclusion of  non-sphericity in
the dynamics  introduces a simple dependence  of the critical collapse
density on the halo mass. The resulting first crossing distribution
yields a better fit to the halo mass functions measured in N-body
simulation (Sheth \& Tormen 2002). However, other modifications to the
original excursion set approach, such as the inclusion of non-radial
degrees of freedom, might also improve the theoretical mass function
(Audit \etal 1997; Del Popolo \& Gambera 1998).  It would thus be 
very desirable to identify additional distinctive predictions  of
the  ellipsoidal collapse model beyond the mass function and bias to
further test this theory.

While earlier numerical studies have not provided any conclusive
evidence for a dependence of halo properties on environment  (Lemson
\& Kauffmann 1999; Percival \etal  2003; Zentner \etal 2005), recent
numerical investigations indicate that, at fixed halo mass, haloes in
dense regions form at (slightly) higher redshift than in low density
environments (Sheth \& Tormen 2004; Avila-Reese 2005; Harker \etal
2005). Using the Millennium Run (Springel  \etal 2005), Gao, Springel
\& White (2005) have convincingly shown that the clustering of haloes
of a fixed mass depends on formation time. This dependence is strong
for haloes with mass less than the typical collapsing mass $M_\star$,
and fades rapidly for $M>M_\star$. Subsequent studies have
demonstrated that many other halo properties, such as spin parameter
or concentration, correlate with the halo assembly history (Maulbetsch
\etal 2006; Wechsler \etal 2006; Zhu \etal 2006; Gao \& White 2007;
Wetzel \etal 2007; Jing, Suto \& Mo 2007). Also, haloes that have
undergone major mergers may be more strongly clustered relative to
other haloes of the same mass (e.g. Furlanetto \& Kamionkowski 2006).
These results call into question the simplest descriptions of
structure formation based on the statistics of random walks (Bond
\etal 1991; Lacey \& Cole 1993; White 1996).  However, relaxing the
assumption of sphericity and/or  sharp $k$-space filtering can
introduce a dependence on environment (Bond \etal 1991; White 1996; 
Sandvik \etal 2007). In the ellipsoidal model, the time required for a 
given overdensity to virialize increases monotonically with the initial
shear (Sheth, Mo \& Tormen 2001). Therefore, as recognised by Wang, Mo
\& Jing (2006), the ellipsoidal dynamics should give an environmental
effect owing to the tidal field generated by the large scale
environment.

In the present paper, we take an analytic approach and assess  whether
the ellipsoidal collapse model can produce an environmental dependence
(also termed 'assembly bias') similar to that seen in N-body
simulations. We start with the statistical dependence that arises in
correlated (Gaussian) initial conditions.  We extend the results of
Doroshkevich (1970) to the joint statistics of the shear tensor. We
derive conditional distributions and quantify the extent to which the
asymmetry of initially triaxial perturbations is sensitive to the
geometry of the large scale environment. Next, we investigate the
dynamical aspect of the environmental dependence using a simplified
model that takes into account the interaction between a triaxial
protohalo and its environment. We find that the tidal force exerted by
the surrounding mass distribution affects the axes collapse and causes
haloes embedded in large overdensities to virialize earlier. A moving
barrier whose shape  depends on the environment density provides a
good description of this environmental effect. This enables us to
apply the EPS formalism in order to estimate  the environmental
dependence of halo properties. Our approach thus is very different
than the multidimensional extension presented in Sandvik \etal (2007).

The paper is organised as follows. Section~\ref{sec:background} briefly
reviews the basic concepts  associated with the ellipsoidal collapse
model. Section~\S\ref{sec:ics} is devoted to the statistical
correlation between the local properties of the shear and the large
scale environment (Appendix~\ref{app:so3} details a delicate step of
the calculation). \S\ref{sec:collapse} investigates the  dynamical
origin of the environmental dependence, focusing on the distribution
of halo formation redshift, large scale bias and alignment of spin
parameter. Non-Markovianess and tidal interactions as potential
sources of environmental dependence are discussed
in~\S\ref{sec:discussion}. A final section summarises our results.

\section{Theoretical background}
\label{sec:background}

We emphasise the role played by the shear in current theories of
structure formation and introduce the basic  definitions and relations
relevant to the study of environmental effects in the ellipsoidal
collapse model.

\subsection{Shear tensor}

The comoving Eulerian position of a particle can be generally
expressed  as a mapping $\vx=\vq+\vvs(\vq,t)$, where $\vq$ is the
Lagrangian (initial)  position and $\vvs$ is the displacement
field. In the Zeldovich approximation (1970), the displacement field
is $\vvs(q,t)=-D(t)\grad\Phi(\vq)$, where  $\Phi(\vq)=\phi(\vq,t)/4\pi
G\rb(t)a^2 D(t)$ is the perturbation potential ($\phi(\vq,t)$ is the
Newtonian gravitational potential), $\rb$ is the average matter
density and $D(t)$ is the linear growth factor (Peebles  1980). The
second derivatives of the perturbation potential define the
deformation tensor (or strain field) ${\rm D}_{ij}=\partial_i
\partial_j\Phi$. For  convenience, we introduce the real, symmetric
tensor
\begin{equation}
\xi_{ij}(\vq)=\frac{1}{\sigma}{\rm D}_{ij}(\vq)=\frac{1}{\sigma}\,
\frac{\partial^2\Phi}{\partial q_i\partial q_j}(\vq)\;,
\label{eq:strain}
\end{equation}
where $\sigma=\sigma(R)$ is the rms variance of density fluctuations
smoothed on scale $R$ (see~\S\ref{sub:joint} below). We will
henceforth refer to $\xi_{ij}$ as the shear tensor. Let
$\lambda_1\geq\lambda_2\geq\lambda_3$ designate the ordered
eigenvalues  of $\xi_{ij}$. An important quantity is the probability
distribution  of the ordered set $(\lambda_1,\lambda_2,\lambda_3)$,
first derived  for Gaussian random fields by Doroshkevich (1970),
\begin{equation}
P(\lambda_1,\lambda_2,\lambda_3)=\frac{15^3}{8\pi\sqrt{5}}\,
\Delta(\lambda)\,e^{-3s_1^2(\lambda)+\frac{15}{2}s_2(\lambda)}\;,
\label{eq:dorosh}
\end{equation}
where
\begin{equation}
\Delta(x)={\rm det}\left(x_i^{N-j}\right)= \prod_{1\leq i<j\leq N}
\left(x_i-x_j\right)
\label{eq:vandermonde}
\end{equation}
is the Vandermonde determinant in the arguments $x_i,i=1,\dots,N$, and
$s_n(x)$ are the elementary symmetric functions of degree $n$ (Weyl
1948).  For three variables $x_1,x_2,x_3$, the first few elementary
functions are
\begin{eqnarray}
s_1(x)\!\!\!\! &=& \!\!\!\! x_1+x_2+x_3 \nonumber \\ s_2(x)\!\!\!\!
&=& \!\!\!\! x_1 x_2+ x_1 x_3+ x_2 x_3 \nonumber \\ s_3(x)\!\!\!\! &=&
\!\!\!\! x_1 x_2 x_3 \;.
\label{eq:esym}
\end{eqnarray}
If $x_1,\dots,x_N$ are the eigenvalues of a matrix $\vxx$, the
functions $s_n(x)$ can be written in terms of the traces of power of
$\vxx$,  $\tr\vxx^k$ with $k,l=0,1,\dots$. For instance,
$s_2(x)=(1/2)\left[\left(\tr\vxx\right)^2-\tr\left(\vxx^2\right)\right]$
etc.

\subsection{Geometry of the initial density field}

As shown by Bond, Kofman \& Pogosyan (1996), the filamentary pattern
seen in N-body simulations (e.g. Park 1990; Bertschinger  \& Gelb
1991; Cen \& Ostriker 1993; Springel \etal 2005 for a recent example)
is a consequence of the initial spatial coherence of the shear tensor.
In this Cosmic Web paradigm,  the correspondence between large scale
structures in the evolved density field and local properties of the
shear tensor in the initial conditions, and the knowledge of the
probability $P(\lambda_1,\lambda_2,\lambda_3)$, allows us estimate the
morphology of the large scale matter distribution.

The geometry of the primeval density field depends on the signature of
the ordered sequence of shear eigenvalues.  If, in a given region, the
largest eigenvalue only is positive ($+--$), there is contraction
along one direction and expansion in the other two so that a pancake
will form. If two eigenvalues are positive while the third one is
negative ($++-$), collapse occurs along two directions and a filament
will form. The probability for these two configurations, $\sim 0.84$,
is much  larger than the probability that all three  eigenvalues are
positive, $P(+++)=0.08$. However, these values depend strongly on the
density  enhancement of  the region under consideration. While
filaments or sheet-like configurations are favoured when $\nu\lsim
1.5$, one  encounters predominantly spherical-like mass concentrations
above $\nu\simeq 1.5$ (e.g. Bernardeau 1994; Pogosyan \etal 1998).
Note that several methods have recently  been proposed to quantify
precisely the geometry and topology of density fields (e.g. Sahni
\etal 1998; Colombi, Pogosyan \& Souradeep 2000; Hanami 2001; Novikov,
Colombi \& Dor\'e 2006; Gleser \etal 2006; Arag\'on-Calvo \etal 2007a).

The sequence in which these structures form is still an open
problem. We briefly note that, in the pancake picture, the collapse
proceeds in the order pancakes\--filaments\--clusters (Lin, Mestel \&
Shu 1965; Zeldovich 1970; Arnold, Shandarin \& Zeldovich 1982).
Supporting evidence comes from several N-body simulations showing
first pancake-like collapse (e.g. Shandarin \etal 1995).  

Identifying the precursors of haloes (proto\-haloes) in the initial
conditions is another unresolved issue, despite the major advance made
in the analysis of (Gaussian) random fields (Doroshkevich 1970; Adler
1981; Peacock \& Heavens 1985; Bardeen \etal 1986; Bond \& Myers
1996). Shandarin \& Klypin (1984) have shown by means of simulations
that massive clusters with $M\bsim 10^{15}\mdh$ are initially located
close to local maxima of the smallest eigenvalue, with $\lambda_3>0$.
On the other hand, recent N-body investigation (Porciani, Dekel \&
Hoffman 2002a,b) indicate  that a large fraction of small mass haloes
$M\leq M_\star$ are rather  associated with primeval configurations of
signature ($++-$). Since, at present, there is no reliable alternative
to the Press-Schechter prescription, we assume that haloes form
out of initially spherical patches of size $R$, so that the relevant
averages are taken over spherical Lagrangian regions. This assumption 
is quite unrealistic in the light of numerical results, but it greatly 
facilitates the calculation. We shall also adopt the usual critical
density criterion issued from the spherical collapse (see below). This
restrictive collapse condition does not guarantee a strictly positive
signature ($+++$). But in the ellipsoidal collapse model, the
formation of a bound object  can occur even if $\lambda_3<0$~:  once
the shortest axis has collapsed,  the nonlinear density causes  the
other two axes to collapse very rapidly (Bond \& Myers 1996).

\subsection{Ellipticity, prolateness and critical collapse density}

The eigenvalues $\lambda_i$ can be equivalently  parametrised in
terms of the shear ellipticity $e$ and prolateness $p$, where
\begin{equation}
e=\frac{\lambda_1-\lambda_3}{2\nu},~~~
p=\frac{\lambda_1-2\lambda_2+\lambda_3}{2\nu}\;.
\label{eq:ep}
\end{equation}
and $\nu=\delta/\sigma=\lambda_1+\lambda_2+\lambda_3$ is the density
contrast of the region under consideration. The ordering constraint
implies that $e\geq 0$ if $\nu>0$ and $e\leq 0$ if $\nu<0$. In all
case, the shear prolateness is $-e\leq p\leq e$. In this
parametrisation, extreme sheet-like (oblate) structures have  $p\lsim
+e$ while extreme filaments (prolate) have $p\bsim -e$.
Doroschkevich's formula can be used to  write down the distribution
$g(e,p|\nu)$ of ellipticity $e$ and  prolateness $p$ for a given
density contrast $\nu$ (e.g. Bardeen \etal 1986),
\begin{equation}
g(e,p|\nu)=\frac{1125}{\sqrt{10\pi}}\, e\left(e^2-p^2\right)\nu^5
e^{-\frac{5}{2}\nu^2\left(3e^2+p^2\right)}\;.
\label{eq:gep1}
\end{equation}
For all $\nu$, the maximum of this distribution occurs at
\begin{equation}
e_m(\nu)=1/(\sqrt{5}\nu),~~~p_m(\nu)=0 \;,
\label{eq:em}
\end{equation}
while the variances of $e$ and $p$ are (for $\nu >0$ only)
\begin{equation}
\la e^2|\nu\ra-\la e|\nu\ra^2=(19\pi-54)/(60\pi\nu^2),
~~\la p^2|\nu\ra=19/(20\nu^2)\;.
\label{eq:e2m}
\end{equation}
Notice that the most probable value $e_m$ is comparable to the mean
ellipticity $\la e|\nu\ra=3/(\sqrt{10\pi}\nu)\approx 1.20 e_m$. It
depends on the smoothing scale $R$ through
$\nu\propto\delta/\sigma(R)$. At fixed $R$, denser regions are more
likely to be spherical than less  dense regions while, at fixed $\nu$,
larger regions are more likely to be spherical than smaller ones.
The scatter in the asymmetry parameters increases strongly 
with decreasing density and/or halo mass.

In the excursion set formalism (Bond \etal 1991; Lacey \& Cole 1993),
the critical collapse density encodes the details of the collapse
dynamics. In the spherical collapse model, the dynamics in a given
cosmological background is governed by a single parameter, namely the
density. A top-hat perturbation  of overdensity $\nu=\dsc/\sigma$
(linearly extrapolated to present epoch) collapses at redshift
$z=0$. The (linear) critical density  threshold $\dsc$ depends on the
cosmology.  We have $\dsc=1.673$ for the cosmological parameters
considered here (Eke, Cole \& Frenk 1996; Navarro, Frenk \& White
1997; Lokas \& Hoffman 2001). On the other hand, in the ellipsoidal
collapse dynamics, the evolution of a perturbation depends on the
values of $e$, $p$ and $\nu$. The critical density threshold $\dec$ is
always larger than the spherical value $\dsc$ and is very sensitive 
to the initial shear (Sheth, Mo \& Tormen 2001).

\section{Environmental effect from the statistics of the initial shear}
\label{sec:ics}

The statistics of the shear tensor  for Gaussian random fields has
been pioneered by Doroshkevich (1970) to study the formation of large
scale structures. In his seminal paper, Doroshkevich calculated the
probability  distribution of the shear eigenvalues and ascertained 
the amount of material being incorporated in a pancake. Later,
Doroshkevich \& Shandarin  (1978) reexamined the formation of
sheet-like structures and derived a distribution function for the
largest eigenvalue of the shear tensor. Recently, Lee \& Shandarin
(1998) computed conditional probability distributions for individual
shear eigenvalues to obtain an analytic approximation to the halo mass
function.

Here, we derive expressions for a number of joint statistics of the
shear tensor upon the assumption of Gaussianity. This enables us to
quantify the importance of the statistical correlation between the
asphericity of triaxial collapsing regions and the shape of their
large scale environment.

\subsection{Analytic considerations}
\label{sub:joint}

We confine our calculation to the case in which the components
$\xi_{ij}(\vx)$ and $\xi_{kl}(\vx')$ are smoothed on different scales,
but the joint distribution is evaluated at a single comoving position
$\vx=\vx'$. This is most relevant to the issues considered in this
paper. The central results of this Section are the joint probability
distribution of shear eigenvalues, eq.~(\ref{eq:gal}), and the
conditional probability for the shear ellipticity and prolateness,
eq.~(\ref{eq:gep2}).

\subsubsection{Spectral parameter}

We begin with the two-point correlation functions of the shear tensor.
The general form of these correlations is given in
Appendix~\S\ref{app:gcase}.  Evaluated at a single comoving position
$\vx$, they take the simple form
\begin{equation}
\la\xi_{ij}(\vx)\xi_{kl}(\vx)\ra=\frac{\gamma}{15}\left(
\delta_{ij}\delta_{kl}+\delta_{ik}\delta_{jl}+\delta_{il}\delta_{jk}
\right)\;,
\label{eq:corxij}
\end{equation}
where $\xi_{ij}$ and $\xi_{kl}$ are smoothed on comoving scale $R_0$
and $R_1$ respectively. The spectral parameter
\begin{equation}
\gamma\equiv \frac{1}{\sigma_0\sigma_1} \int_0^\infty\!\!\dd\ln
k\,\dkk\, \hw(R_0,k)\hw(R_1,k)\;,
\label{eq:xi01}
\end{equation}
$0\leq\gamma\leq 1$, is a measure of the correlation between  these
scales.  Here, $\dkk\equiv k^3 P_\delta(k)/2\pi^2$ is the
dimensionless, linear density power spectrum (Peebles  1980) and
$\sigma_0$ and $\sigma_1$ are the rms variances of density
fluctuations smoothed on scale $R_0$ and $R_1$, respectively. $R_i$ is
the  comoving characteristic scale of the spherically symmetric window
function $\hw(R_i,k)$. Many choices are possible for this filtering
function.  We will  adopt a top-hat filter  throughout this paper. The
top-hat smoothing radius $R_i$ defines a mass scale $M_i=(4\pi/3)\rb
R_i^3$ so that, for a given  power spectrum, $\sigma_i$, $M_i$ and
$R_i$ are equivalent variables.  Note also that, instead of writing
explicitely the smoothing radius, we will use subscripts to
distinguish quantities at different smoothing lengths.  We will
reserve the subscript 1 for haloes and the subscript 0 for the
environment, assumed uncollapsed at $z=0$.

\begin{figure}
\center \resizebox{0.45\textwidth}{!}{\includegraphics{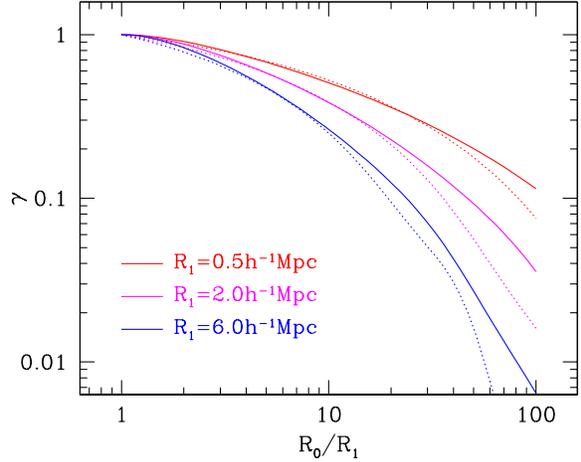}}
\caption{The spectral parameter $\gamma$ as a function of the ratio
$R_0/R_1$  for  three different smoothing lengths $R_1=0.5$, 2 and
6$\hmpc$ (solid curves  from top to bottom). These values correspond
to a mass scale $M_1=3.3\times 10^{10}$,  $2.2\times 10^{12}$ and
$5.8\times 10^{13}\mdh$, respectively. The dotted curves show the
scaling $\gamma\propto (R_1/R_0)^{-(3+n_{\rm eff})/2}$ (see text).  In
all cases, $\gamma$ is large even on scales $R_0\gg R_1$.}
\label{fig:gam}
\end{figure}

Fig.~\ref{fig:gam} shows the correlation strength $\gamma$ as a
function of $R_0$ for  three different values of $R_1$~: 0.5, 2 and
6$\hmpc$ (curves from top to bottom).  The curves have been computed
for a $\Lambda$CDM model of spectral index $n_s=0.96$ and
normalisation $\sigma_8=0.83$ using the fitting formulae of Eisenstein
\& Hu (1999). This choice is consistent with the  constraints inferred
from the latest CMB measurements (WMAP3, see Spergel \etal  2006).
For the special case of a power-law spectrum $P_\delta(k)\propto k^n$,
the parameter $\gamma$ scales with the smoothing lengths $R_0$ and
$R_1$ as
\begin{equation}
\gamma\propto\left(\frac{R_1}{R_0}\right)^{(3+n)/2}\;.
\label{eq:approx}
\end{equation}
Strictly speaking, this expression is valid for a power-law spectrum
only.  However, it provides a reasonable approximation in the
$\Lambda$CDM  cosmology considered here if the spectral index $n$ is
replaced by an  effective index $n_{\rm eff}(k)\equiv\dd\ln
P_\delta(k) /\dd\ln k$ evaluated on  scale $k\sim 1/R_0$. The scaling
$\gamma\propto (R_1/R_0)^{-(3+n_{\rm eff})/2}$  is plotted as dotted
curves in Fig.~\ref{fig:gam}. Recall that the spectral index is close
to  $n_{\rm eff}\sim -2$ on comoving scales $1-10\hmpc$.

\subsubsection{Joint statistics of the shear tensor}

Owing to the symmetry of $\xi_{ij}$, only six components are
independent. We adopt the notation of Bardeen \etal (1986) and label
them by $\xi_A$, where the components $A=1,\dots,6$ of the
six-dimensional vector refer to the components $ij=11,22,33,12,13,23$
of the tensor.  The joint probability distribution
$P\left(\xi_0,\xi_1\right)$ of the  shear tensor $\xi_0$ and $\xi_1$,
smoothed on scale $R_0$ and $R_1$  respectively, is given by a
multivariate Gaussian whose covariance  matrix $\vmm$ has 12
dimensions.  This $12 \times 12$ matrix may be partitioned into four
$6\times 6$  block matrices, $\vmm_1=\la\ve_1\ve_1^\top\ra$ in the top
left corner,  $\vmm_2=\la\ve_0\ve_0^\top\ra$ in the bottom right
corner,  $\vbb=\la\ve_0\ve_1^\top\ra$ in the bottom left corner and
its transpose  $\vbb^\top$ in the top right corner.  Following Bardeen
\etal (1986), we transform the six dimensions
$\left\{\xi_{0,A},\xi_{1,A},A=1,2,3\right\}$ to a new set of variables
$\left\{u_k,v_k,w_k,k=0,1\right\}$, where
\begin{eqnarray}
&& u_k = \xi_{k,1}+\xi_{k,2}+\xi_{k,3} \nonumber \\
&& v_k = \frac{1}{2}\left(\xi_{k,1}-\xi_{k,3}\right) \nonumber \\ 
&& w_k = \frac{1}{2}\left(\xi_{k,1}-2\,\xi_{k,2}+\xi_{k,3}\right) \;.
\label{eq:newset}
\end{eqnarray}
With these definitions,
\begin{eqnarray}
&& \la u_k^2\ra=1,~~\la v_k^2\ra=\frac{1}{15},~~\la
w_k^2\ra=\frac{1}{5} \nonumber \\ && \la u_0
u_1\ra=\gamma,~~\la v_0 v_1\ra=\frac{\gamma}{15}, ~~\la w_0
w_1\ra=\frac{\gamma}{5} \;,
\label{eq:cov1}
\end{eqnarray}
The other correlations are zero. For the six remaining components
$\left\{\xi_{0,A},\xi_{1,A},A=4,5,6\right\}$, the correlations
functions are
\begin{eqnarray}
&& \la\xi_{0,A}\xi_{0,B}\ra=\la\xi_{0,A}\xi_{0,B}\ra=
\frac{1}{15}\,\delta_{AB} \nonumber \\ &&
\la\xi_{0,A}\xi_{1,B}\ra=\frac{\gamma}{15}\,\delta_{AB} \;.
\label{eq:cov2}
\end{eqnarray}
All the cross-correlations between $u_k$, $v_k$, $w_k$ and
$\xi_{0,A}$,  $\xi_{1,A}$ vanish. The block matrices $\vmm_1$,
$\vmm_2$ and $\vbb$ are  diagonal in the basis introduced above,
\begin{equation}
\vmm_1=\vmm_2=\left(\begin{array}{cc}\vcc & 0 \\  0 &
\vii/15\end{array}\right),~~~ \vbb=\left(\begin{array}{cc}\gamma\vcc &
0 \\  0 & \gamma\vii/15\end{array}\right) \;,
\end{equation} 
where $\vcc={\rm diag}(1,1/15,1/5)$ and $\vii$ is the $3\times 3$
identity matrix. The quadratic form which appears in the joint
probability distribution
\begin{equation}
P(\xi_0,\xi_1)\dd\xi_0\dd\xi_1=\frac{1}{\left(2\pi\right)^6
  |\det\vmm|^{1/2}}\,e^{-Q(\xi_0,\xi_1)}\dd\xi_0\dd\xi_1 \;,
\label{eq:pdf1}
\end{equation}
where $\det\vmm$ is the determinant of the covariance matrix $\vmm$,
can be  computed easily using Schur's identities. The result may be
expressed  in terms of the elementary symmetric functions
(\ref{eq:esym}) or, equivalently, in terms of traces,
\begin{eqnarray}
\lefteqn{Q\left(\xi_0,\xi_1\right)=\frac{3}{4\left(1-\gamma^2\right)}
\left\{5\left[\tr\left(\xi_0^2\right)+\tr\left(\xi_1^2\right)-
2\gamma\tr\left(\xi_0\xi_1\right)\right]\right.} \nonumber \\ &&
-\left. \left[\left(\tr\xi_0\right)^2+
\left(\tr\xi_1\right)^2-2\gamma\left(\tr\xi_0\right)\left(\tr\xi_1\right)
\right]\right\}\;,
\label{eq:Q}
\end{eqnarray}
and the square-root of the determinant is given by
$|\det\vmm|^{1/2}=(20/15^6)(1-\gamma^2)^3$. The results are described
by the correlation strength $\gamma$ solely. The invariance under
rotation $P\left(\xi_0,\xi_1\right)=
P\left(\vrr\xi_0\vrr^\top,\vrr\xi_1\vrr^\top\right)$, where $\vrr$  is
a real orthogonal symmetric $3\times 3$ matrix, requires that $P$ be a
symmetric function of the eigenvalues, and thus a  function of
$\tr\left(\xi_0^k\xi_1^l\right)$, $k,l=0,1,\dots$,  regardless the
statistical properties of $\xi_{ij}$. The  expression (\ref{eq:Q})
follows from our assumption of Gaussianity.  Note also that no
assumptions have been made so far about the coordinates.

\subsubsection{Joint distribution of the eigenvalues}
\label{subsub:eigen}

Lee \& Shandarin (1998, see their Appendix B) have computed the joint
probability distribution of the eigenvalues of the deformation tensor
for a sharp $k$-space filter. However, they assume that both principal
axis frames are aligned, which is not true in general.

To obtain the joint probability distribution of  the ordered
eigenvalues of the shear tensor, we choose a  coordinate system such
that the coordinate axes are aligned with the  principal axes of
$\xi_0$. Let $\alpha$ and $\lambda$ be the diagonal  matrices
consisting of the three ordered eigenvalues
$\alpha_1\geq\alpha_2\geq\alpha_3$ and
$\lambda_1\geq\lambda_2\geq\lambda_3$ of the deformation tensors
$\xi_0$ and $\xi_1$, respectively. The principal axis are labelled
according to this ordering. With this choice of coordinate,
$\xi_0=\alpha$ and $\xi_1=\vrr\lambda\vrr^\top$, where $\vrr$ is an
orthogonal matrix that defines the orientation of the eigenvectors of
$\xi_1$ relative to those of $\xi_0$. To preserve the orientation  of
the principal axis frames, we further impose the condition that the
determinant of $\vrr$ must be +1. Namely, $\vrr$ belongs to the
special  orthogonal group SO(3).  The properties of the trace imply
that $\tr\,\xi_1=\tr\,\lambda$ and
$\tr\left(\xi_1^2\right)=\tr\left(\lambda^2\right)$. We note, however,
that the term
$\tr\left(\xi_0\xi_1\right)=\tr\left(\vrr\lambda\vrr^\top\!\alpha\right)$
depends on the rotation matrix.

The joint probability distribution $P(\alpha,\lambda)$ is obtained by
integrating over the rotations that define the orientations of the
orthonormal eigenvectors of $\xi_0$ and $\xi_1$.  The volume measure
$\dd\xi$ for the space of real $3\times 3$ symmetric  matrices can be
expressed in terms of the non-increasing sequence of  eigenvalues
$t_i$ as
\begin{equation}
\dd\xi=8\pi^2\,\Delta\left(t\right)\dd^3 t\,\dd\vrr\;.
\label{eq:vol1}
\end{equation}
Here, $\dd\vrr$ is the invariant measure on the group SO(3) normalised
to $\int\dd\vrr=1$, $\Delta(t)$ is the Vandermonde determinant (eq.
\ref{eq:vandermonde}) and $\dd^3 t=\dd t_1\dd t_2\dd t_3$. Since the
quadratic form $Q$ depends only on the relative orientation of the two
orthonormal triads, we can immediately integrate over one of the SO(3)
manifolds.  The relevant volume is $8\pi^2/4=2\pi^2$. The factor 4
comes from not  caring whether the rotated axis points in the positive
or negative  direction (see e.g Bardeen \etal 1986).  The essential
problem is the calculation of the integral over the rotations that
define relative, distinct triad orientations.
Appendix~\S\ref{app:so3} shows that the integral over the second SO(3)
manifold can be cast into the form
\begin{equation}
\frac{1}{4}\int_{\rm SO(3)}\!\!\!\!\!\!\dd\vrr\,
\exp\left[\beta\,\tr\left(\vrr\lambda\vrr^\top\!\alpha\right)\right]=
\frac{e^{\beta\,\epsi_+}}{4}\,
W\left(\beta\,\epsi_-,\epsi_\alpha,\epsi_\lambda\right)\,
\label{eq:iso3}
\end{equation}
where the function W depends on $\gamma$ through the parameter
$\beta(\gamma)=(15/2)\gamma/(1-\gamma^2)$. The 
four independent variables $\epsi_+$, $\epsi_-$, $\epsi_\alpha$ and
$\epsi_\lambda$ are combinations of the six eigenvalues of $\alpha$
and $\lambda$ (Wei \& Eichinger 1990). We choose $\epsi_+=(1/3)
\,\tr\alpha\,\tr\lambda$, $\epsi_-=(1/3)\,\tr\tilde{\alpha}\,
\tr\tilde{\lambda}$,
$\epsi_\alpha=(\alpha_1-\alpha_2)/\tr\tilde{\alpha}$,
$\epsi_\lambda=(\lambda_1-\lambda_2)/\tr\tilde{\lambda}$, where
$\tr\tilde{\alpha}=\tr\alpha-3\alpha_3$ and
$\tr\tilde{\lambda}=\tr\lambda-3\lambda_3$. With this parametrisation,
$-\infty\leq\epsi_+\leq\infty$, $\epsi_-\geq 0$ and
$0\leq\epsi_\alpha,\epsi_\lambda\leq 1$. The function
$W(\beta\,\epsi_-,\epsi_\alpha,\epsi_\lambda)$ can be written 
down as a double integral (eq.~\ref{eq:wfunc}).  We have found that a
fifth-order expansion about $\beta\epsi_-=0$  (eq.~\ref{eq:iwf}) is
accurate to within 2 per cent in the range  $0\leq\beta\epsi_-\lsim
1.5$. We use this truncated series in the computation  of the group
integral (\ref{eq:iso3}). The joint probability distribution
$P(\alpha,\lambda)\dd^3\!\alpha\,\dd^3\!\lambda$ may now be
formulated as
\begin{eqnarray}
P(\alpha,\lambda)\dd^3\!\alpha\,\dd^3\!\lambda \!\!\!\! &=& \!\!\!\!
\frac{15^6}{320\pi^2}\left(1-\gamma^2\right)^{-3}
W(\beta\epsi_-,\epsi_\alpha,\epsi_\lambda) \nonumber \\  && \times
e^{-Q_{01}+\beta\epsi_+}\,
\Delta(\alpha)\Delta(\lambda)\,\dd^3\!\alpha\,\dd^3\!\lambda \;,
\label{eq:pal}
\end{eqnarray}
where the quadratic form $Q_{01}$ is a function of $\alpha$ and
$\lambda$,
\begin{eqnarray}
Q_{01}\!\left(\alpha,\lambda\right) \!\!\!\! &=& \!\!\!\!
\frac{3}{4\left(1-\gamma^2\right)}
\left\{5\left[\tr\left(\alpha^2\right)+\tr\left(\lambda^2\right)\right]
\right. \nonumber \\ && -\left. \left[\left(\tr\alpha\right)^2+
\left(\tr\lambda\right)^2-2\gamma\left(\tr\alpha\right)
\left(\tr\lambda\right)\right]\right\} \;.
\label{eq:Q01}
\end{eqnarray} 
Note that, in the limit $\gamma\ll 1$, $\beta\rightarrow 0$ and the
joint probability distribution $P(\alpha,\lambda)$ tends, as it should
be, toward the product of the individual one-point probability
distribution eq.~(\ref{eq:dorosh}.  Using Bayes' theorem, we can
easily derive the distribution of eigenvalues $\lambda_i$ given the
values of $\alpha_i$. This conditional probability may be written as
\begin{eqnarray}
\lefteqn{P(\lambda|\alpha)=\frac{15^3}{8\pi\sqrt{5}}
\left(1-\gamma^2\right)^{-3}W(\beta\epsi_-,\epsi_\alpha,\epsi_\lambda)\,
e^{\beta\epsi_+}\Delta(\lambda)} \\ &&\times \exp\left[-\frac{
15\gamma^2\tr\left(\alpha^2\right)+15\tr\left(\lambda^2\right)
-3\left(\gamma\tr\alpha-\tr\lambda\right)^2}
{4\left(1-\gamma^2\right)}\right] \nonumber \;.
\label{eq:gal}
\end{eqnarray}
It can be verified by direct numerical integration that the
probability  distribution (\ref{eq:gal}) is normalised to unity (we
have used the multidimensional integrator {\small DCUHRE} described in
Berntsen,  Espelid \& Genz 1991).

\begin{figure}
\center \resizebox{0.45\textwidth}{!}{\includegraphics{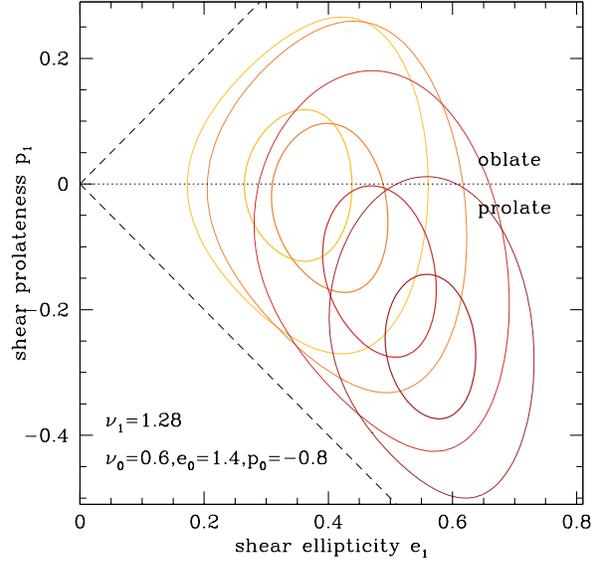}}
\caption{The 20 and 68 per cent contours of the conditional
probability distribution $g(e_1,p_1|\nu_1,0)$ for several values of
the correlation  strength, $\gamma=0$, 0.4, 0.6 and 0.8 (from top to
bottom). The values of $(e_0,p_0,\nu_0)$ are typical of a (prolate)
filament. The dashed  lines indicate the boundary of the domain
$|p|\geq e$.}
\label{fig:gep1}
\end{figure}

\subsubsection{Joint distribution of the shear ellipticity and prolateness}

The results of Section~\S\ref{subsub:eigen} can be conveniently
expressed in terms of the density contrast $\nu$, shear ellipticity
$e$ and shear  prolateness $p$. The joint probability distribution for
these variables,  $g\left(e_1,p_1,\nu_1,e_0,p_0,\nu_0\right)$, is
readily obtained from  the following coordinate transformation,
\begin{eqnarray}
&&\alpha_1=\frac{\nu_0}{3}\left(1+3e_0+p_0\right)~~~
\lambda_1=\frac{\nu_1}{3}\left(1+3e_1+p_1\right) \nonumber \\
&&\alpha_2=\frac{\nu_0}{3}\left(1-2p_0\right)\hspace{10mm}
\lambda_2=\frac{\nu_1}{3}\left(1-2p_1\right) \nonumber \\
&&\alpha_3=\frac{\nu_0}{3}\left(1-3e_0+p_0\right)~~~
\lambda_3=\frac{\nu_1}{3}\left(1-3e_1+p_1\right) \;.
\label{eq:ei}
\end{eqnarray}
We have, for example, $\tr(\alpha)=\nu_0$,
$\dd^3\alpha=(2/3)\nu_0^2\,\dd\nu_0\dd e_0\dd p_0$  and
$\Delta(\alpha) = 2\nu_0^3\, e_0\left(e_0^2-p_0^2\right)$. With this
parametrisation, the quadratic form $Q_{01}$ becomes
\begin{eqnarray}
Q_{01} \!\!\!\! &=& \!\!\!\! \frac{5}{2\left(1-\gamma^2\right)}
\left[\nu_0^2\left(3e_0^2+p_0^2\right)+\nu_1^2\left(3e_1^2+p_1^2
\right)+\gamma\nu_0\nu_1\right] \nonumber \\ &&
+\frac{1}{2}\left[\frac{\left(\nu_1-\gamma\nu_0\right)^2}
{1-\gamma^2}+\nu_0^2\right] \;.
\label{eq:Q1ep}
\end{eqnarray}
The other variables are simply $\epsi_+=(1/3)\nu_0\nu_1$,  $\epsi_- =
(1/3) \nu_0\nu_1 \left(3e_0-p_0\right)\left(3e_1-p_1\right)$,
$\epsi_\alpha=\left(e_0+p_0\right)/\left(3e_0-p_0\right)$ and
$\epsi_\lambda=\left(e_1+p_1\right)/\left(3e_1-p_1\right)$.

\begin{figure*}
\center 
\resizebox{0.45\textwidth}{!}{\includegraphics{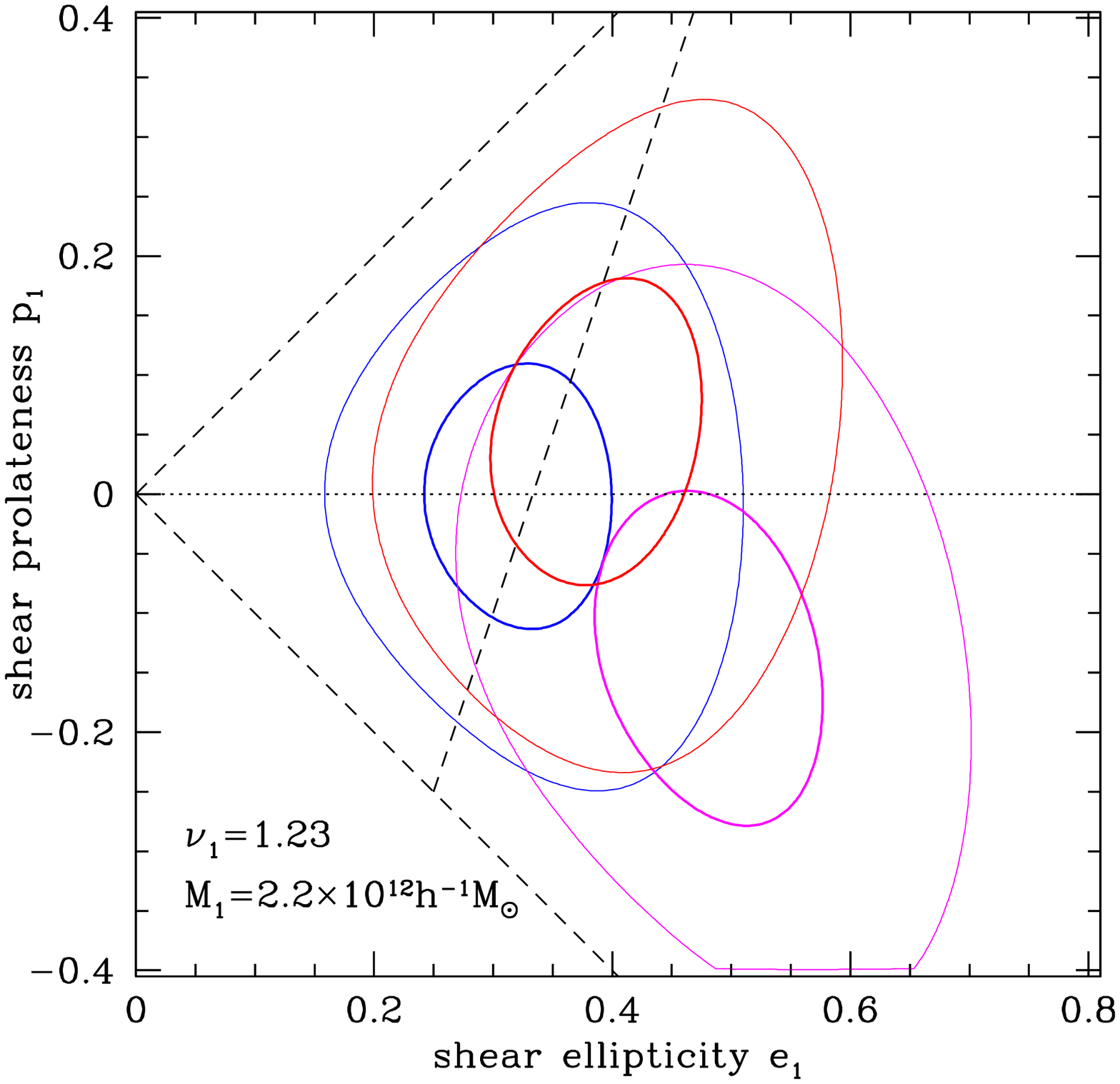}}
\resizebox{0.45\textwidth}{!}{\includegraphics{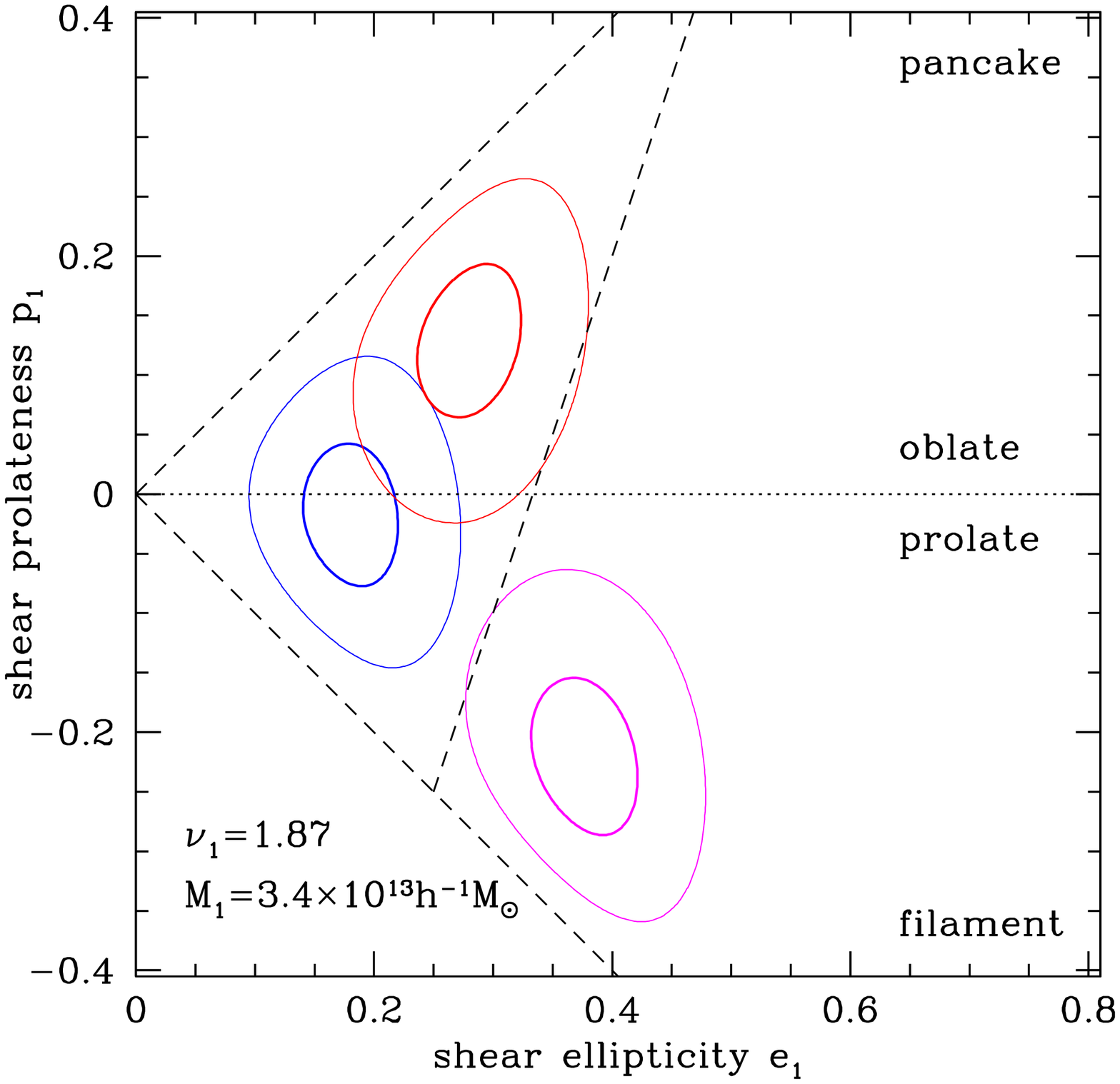}}
\caption{The 20 and 68 per cent percentiles of $g(e_1,p_1|\nu_1,0)$
for two different halo mass $M_1=2.2\times 10^{12}$ (left panel) and
$M_1=3.4\times 10^{13}\mdh$ (right panel). These mass scales
correspond to a smoothing radius $R_1=2$ and $5\hmpc$,
respectively. On large scale, the shear is smoothed on a fixed radius
$R_0=10\hmpc$ ($M_0=2.7\times 10^{14}\mdh$). The shape of the large
scale region  is either a proto-cluster, the precursor of a filament
or a sheet-like structure, characterised by
$(e_0,p_0,\nu_0)=(0.3,-0.1,1)$, $(1.5,-1.3,0.5)$  and $(1.3,1.1,0.4)$,
respectively. The dashed lines indicate the boundary of the domain
$|p_1|\geq e_1$. The interior of the triangle  bounded by
$(e_1,p_1)=(0,0)$, $(\frac{1}{4},-\frac{1}{4})$ and
$(\frac{1}{2},\frac{1}{2})$ shows the region where $\lambda_3>0$.  For
a given mass $M_1$, proto\-haloes which collapse within filament-like
or sheet-like structures are, on average, initially more asymmetric
than those which will form  in spherical-like environment. The average
asphericity and the scatter increase with decreasing halo  mass.}
\label{fig:gep2}
\end{figure*}

Let us introduce the notational shorthands $0\equiv (e_0,p_0,\nu_0)$
and $1\equiv (e_1,p_1,\nu_1)$. We wish now to calculate the
conditional  probability $g(e_1,p_1|\nu_1,0)$  of having an
ellipticity $e_1$ and prolateness $p_1$ given a density  $\nu_1$ on
scale $R_1$, and given the values $(e_0,p_0,\nu_0)$ on  scale
$R_0$. This probability will be useful to estimate the effect  of the
large scale environment on the primeval distribution of shear
ellipticity and prolateness. Bayes' theorem implies that
\begin{equation}
g\left(e_1,p_1|\nu_1,0\right)=\frac{g\left(0,1\right)}
{g\left(\nu_1,0\right)}\;.
\label{eq:gep11}
\end{equation}
Since $\la \nu_1\xi_{0,A}\ra=\gamma/3$ if $A=1,2,3$ and the other
cross-correlations are zero, the calculation of the denominator is
straightforward. We have $g(\nu_1,0)=g(\nu_1|\nu_0)\,g(0)$, where
\begin{equation}
g(\nu_1|\nu_0)=\frac{1}{\sqrt{2\pi}}\left(1-\gamma^2\right)^{-1/2}
\exp\left[-\frac{1}{2}\frac{\left(\nu_1-\gamma\nu_0\right)^2}
{\left(1-\gamma^2\right)}\right]\;,
\label{eq:gep12}
\end{equation}
as the density contrast $\nu$ is independent of the asymmetry
parameters $e$ and $p$. With these informations, the conditional
probability distribution can be expressed as
\begin{equation}
g(e_1,p_2|\nu_1,0)=b(0,1)\,g(e_1,p_1|\nu_1)\;,
\label{eq:gep2}
\end{equation}
where
\begin{equation}
g(e_1,p_1|\nu_1)=\frac{1125}{\sqrt{10\pi}}\,\nu_1^5 e_1 
\left(e_1^2-p_1^2\right)e^{-\frac{5}{2}\nu_1^2\left(3 e_1^2+p_1^2\right)}
\label{eq:gepnoenv}
\end{equation}
is the distribution without the environmental constraint (see, 
e.g., Sheth, Mo \& Tormen 2001), and
\begin{eqnarray}
\lefteqn{b(0,1)=\left(1-\gamma^2\right)^{-5/2}W\left(\beta\epsi_-,
\epsi_\alpha,\epsi_\lambda\right)} \\
&& \times \exp\left\{-\frac{5\gamma^2}{2\left(1-\gamma^2\right)}\left[
\nu_0^2\left(3 e_0^2+p_0^2\right)+\nu_1^2\left(3 e_1^2+p_1^2\right)
\right]\right\}\nonumber
\label{eq:gepbias}
\end{eqnarray}
is a correction factor which results from the constraint ``0''.
As expected, in the limit where the correlation becomes weak, the
conditional distribution $g(e_1,p_1|\nu_1,0)$ reduces to the
unconditional distribution, eq.~(\ref{eq:dorosh}), derived by
Doroshkevich. It vanishes outside the domain defined by $|p_1|\leq
e_1$. We  stress that these expressions are valid for any window
function. In the particular case of a sharp $k$-space filter 
however, the spectral parameter is simply $\gamma=\sigma_0/\sigma_1$.

The  distribution (\ref{eq:gep2}) is sufficient to estimate the
magnitude of the environmental effect which arises from the statistics
of the initial shear field.

\subsection{Statistical correlation between proto\-halo and 
environment in Gaussian initial conditions}
\label{sub:edics}

We illustrate how the local characteristics of the shear tensor
smoothed on the scale of collapsing haloes depend on the initial
geometry of their large scale environment.

\subsubsection{Shear ellipticity and prolateness}

First, it is worthwhile studying how the conditional probability
distribution changes with the  correlation strength $\gamma$. To this
purpose, we evaluate the probability~(\ref{eq:gep2}) for a density
$\delta_1=2$, a value sufficiently large so that the small scale
overdensity collapses at moderate to high redshift regardless of the
asymmetry parameters.  For a filtering scale $R_1=2\hmpc$, i.e. a halo
mass $M_1\simeq 2\times 10^{12}\mdh$, this corresponds to a density
contrast $\nu_1=1.28$.  On large scale  $R_0>R_1$, we take the shear
ellipticity, prolateness and density to be
$(e_0,p_0,\nu_0)=(1.4,-0.8,0.6)$. These values are appropriate for   a
filament-like configuration.   In Fig.~\ref{fig:gep1}, the 20 and 68
percentiles of the conditional probability  distribution
$g(e_1,p_1|\nu_1,0)$ are plotted for $\gamma=0$, 0.4, 0.6 and 0.8
(contours from top to bottom). A larger value of $\gamma$ increases
the asymmetry of the distribution and sharpens its maximum. Recall
that, at fixed value of $R_1$, the correlation strength $\gamma$
decreases with increasing $R_0$ (see Fig.~\ref{fig:gam}). For
instance, with $R_1=2\hmpc$ and a  reasonable environment size
$R_0\sim 10\hmpc$, we have  $\gamma\sim 0.6$. The resultant
distribution is displayed third from  the top. In this case, the most
probable  values of the shear ellipticity and prolateness are
$(e_m,p_m)=(0.48,-0.15)$, significantly different than those in the
limit $\gamma\rightarrow 0$, namely $(0.35,0)$. Note that the top-hat
smoothing artificially reduces the asphericity of the large scale
environment. Therefore, the values of $e_m$ and $p_m$ certainly
underestimate the asymmetry that could be measured from N-body
simulations (Bardeen \etal 1986). Finally,  unless otherwise stated,
we shall adopt $R_0=10\hmpc$ ($M_0=2.7\times 10^{14}\mdh$)  as the
environment radius in the  remaining of this paper.

Fig.~\ref{fig:gep2} further illustrates the environmental dependence
which arises from the statistical properties of the shear tensor.
Contours are plotted for three different configuration shapes of the
large scale environment. The halo mass is now $M_1=2.2\times 10^{12}$
(left panel) and  $M_1=3.4\times 10^{13}\mdh$ (right panel), and the
resultant correlation coefficient is $\gamma=0.58$ and 0.85,
respectively. Also, we choose $\nu_1$ so that $\delta_1=2$ in all
cases.  The configuration shape of the large scale region is either a
proto-cluster of signature ($+++$), the  precursor of a filament
$(++-)$, or a sheet-like   structure $(+--)$. Clearly, at fixed halo
mass $M_1$, the proto\-haloes that collapse within the pancake or the
filament are initially more asymmetric than those that will form in
the spherical-like region.  Furthermore,  the asymmetry increases
noticeably with decreasing halo mass. For  $M_1=2.2\times
10^{12}\mdh$, the most  probable values of $e_1$ and $p_1$ are
$(e_m,p_m)=(0.32,0)$, $(0.48,-0.14)$ and $(0.39,0.05)$ respectively,
whereas, for $M_1=3.4\times 10^{13}\mdh$, we find
$(e_m,p_m)=(0.17,-0.02)$, $(0.37,-0.22)$ and $(0.28,0.14)$.  In the
high mass halo case, the conditional distributions exhibit much less
scatter  as a result of the larger values of $\gamma$ and $\nu_1$.

Since the density contrast is independent of the shear ellipticity and
prolateness, the probability  of finding a virialized halo in a given
environment of initial shape  $(e_0,p_0,\nu_0)$ is  modulated by
$g(\nu_1|\nu_0)$, which is very  sensitive to $\gamma$. For the low
mass halo considered here, this  probability is 0.38, 0.31 and 0.17
for the  cluster, filament and  pancake configurations,
respectively. For the  massive proto\-halo  however, $g(\nu_1|\nu_0)$
is significantly nonzero (0.10) for the spherical-like environment
solely. Thus, low mass haloes form in the mildly overdense structures
of the primeval density field, while the most massive collapse mainly
in the densest, weakly aspherical regions. This is a distinctive
feature of hierarchical formation models (e.g. Kaiser 1984; Mo \&
White 1996) that adds to decrease the scatter in the relation between
the environment and the local properties of the shear with increasing
halo mass. This statistical effect may provide an explanation for the
strong environmental dependence of low mass haloes.

\begin{figure}
\center \resizebox{0.45\textwidth}{!}{\includegraphics{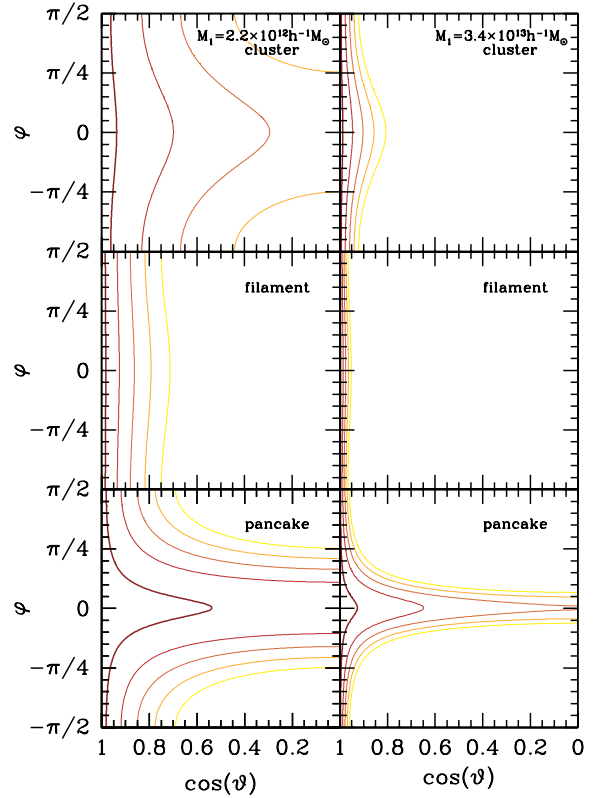}}
\caption{Contours of constant probability $P(\vartheta,\varphi|0,1)$
for the cluster,  filament and pancake-like configurations described
in the text. They  are shown for a halo mass $M_1=2.2\times 10^{12}$
(left panels) and  $3.4\times 10^{13}\mdh$ (right panels). The
asymmetry parameters $e_1$ and $p_1$ assume their most probable
value. Levels of contours decrease by a factor of 2. $\vartheta$  is
the angle between the minor axes. In all cases, the probability  is
highest along the vertical line $\cos\vartheta=1$. The alignment  is
strongest in the case of the filament, whose minor axis coincides
with the axis of symmetry.}
\label{fig:pso3}
\end{figure}

\subsubsection{Alignment of principal axis frames}

It is fairly straightforward to derive probability  distributions for
the relative orientation of the principal axis frame from the results
of \S\ref{sub:joint}.   Appendix~\ref{app:so3} provides the details of
the calculation.  Briefly, we parametrise the rotation matrix $\vrr$
in terms  of the Euler angles $0\leq \varphi,\psi\leq 2\pi$ and
$0\leq\vartheta\leq\pi$. We adopt the ZXZ convention so  that
$\vartheta$ is the angle between the two minor (third) axes. The trace
$\tr\left(\vrr\lambda\vrr^\top\!\alpha\right)$ is then decomposed into
a sum of rotation matrices  ${\cal
D}^l_{_{m_1,m_2}}\!\left(\varphi,\vartheta,\psi\right)$.  The integral
over the  variable $\psi$ can be performed easily and yields the
conditional  probability $P(\vartheta,\varphi|0,1)$ given the values
of $e_0,e_1$   etc. In Fig.~\ref{fig:pso3}, contours of constant
$P(\vartheta,\varphi|0,1)$ are plotted for the cluster, filament and
pancake-like configurations described above. They are shown for the
halo mass $M_1=2.2\times 10^{12}$ (left panels) and $M_1=3.4\times
10^{13}\mdh$ (right panels) considered in Fig.~\ref{fig:gep2}. 
In all cases, the asymmetry parameters
$e_1$ and $p_1$ assume their most probable values. Unsurprisingly, the
alignment between the minor axes of the shear smoothed on scale $R_0$
and $R_1$ is weaker for the low mass halo as a result of the smaller
values of $\gamma$ and $\nu_1$.  At fixed halo  mass however, the
alignment is substantially stronger in the case of the  filamentary
structure (middle panels). This mostly follows from the  fact that,
for a configuration shape with one or two positive  eigenvalues, the
alignment is strongest along the axis of symmetry,  which coincides
with the minor axis for a filamentary structure.  Conversely, the
alignment between the major axes is much stronger  in the pancake
configuration.  These results show  that the principal axis frame of
the tidal tensor smoothed on the proto\-halo mass scale cannot be
assumed independent of that defined by the environment. This is
especially true when the  large scale configuration shape is highly
asymmetric. Accounting for this correlation has a large impact on the
probability  distribution of the spin parameter
(see~\S\ref{subsub:spin}).

\vspace{6pt}

To summarise, a correlation between the local properties of the shear
and the configuration shape of the environment is expected in Gaussian
initial conditions because fluctuations on different scales are
correlated.  Our results nicely demonstrate the large  magnitude of
this statistical effect. At fixed halo mass, the shear of a
perturbation that collapse  within filaments or pancakes is on average
more asymmetric than in  spherical regions. The principal  axis frame
tends to be aligned along the axis of symmetry of the external mass
distribution. Furthermore, the scatter in the relation between the
primeval shear  field and the geometry of the large scale environment
increases strongly with decreasing halo mass. This will almost surely
have a significant impact on the properties of virialized haloes in
the ellipsoidal collapse model since the critical density threshold
$\dec$ depends strongly on the initial values of $e_1$ and $p_1$. We
will further discuss the importance of this statistical correlation in
the next Section.

\section{Environmental dependence in the ellipsoidal collapse dynamics}
\label{sec:collapse}

We now concentrate on the dynamical origin of the environmental
dependence in the ellipsoidal collapse model.  We employ a simplified
model based on the collapse of homogeneous ellipsoids that takes into
account the interaction between a collapsing halo and its
surroundings. To anticipate the results of this Section, we find that
haloes residing in large overdensities virialize earlier. The
environment density is the key parameter in determining the
virialization redshift.    We incorporate this dynamical effect into
the excursion set formalism  by means of a collapse barrier whose
height depends on the environment density. This approach greatly
simplifies the calculation of the  environmental dependence of halo
properties. It also predicts a clear correlation between formation
redshift, large scale bias   and environment.

\subsection{Homogeneous ellipsoidal dynamics}
\label{sub:edyn}

\subsubsection{Equation of motion}

We consider the collapse of a triaxial perturbation embedded in a
uniform background assumed to be a $\Lambda$CDM cosmology. We neglect
the influence of nonlinear substructures on the gravitational
evolution. The initial (Lagrangian) volume occupied by this  overdense
fluctuation is a (uniform) sphere of comoving radius $R_1$.

Following Peebles  (1980) and Eisenstein \& Loeb (1995), the (proper)
position of any point interior to the ellipsoid is conveniently
described  as $r^\alpha=\vaa^{\alpha\beta}q^\beta$, where
$\vq^\top\vq\leq 1$ and repeated indices are summed. The matrix $\vaa$
is a function of time solely. At all time, the equation defining the
outer shell of the ellipsoid is $\vr^\top
\left(\vaa\vaa^\top\right)^{-1}\vr=1$. The principal axis lengths
$\left\{A_k,k=1,2,3\right\}$  and directions of the ellipsoid can be
found by diagonalizing   $\vaa\vaa^\top=\vqq\tilde{\vaa}\vqq^\top$,
where $\vqq$ is orthogonal  and $\tilde{\vaa}$ is a positive definite
diagonal matrix. In this model,  the potential is quadratic in the
coordinates, $\Phi(\vr)=1/2\,\Phi^{\alpha\beta}r^\alpha r^\beta$, so
that the external and internal forces preserve the homogeneity of the
proto\-halo at all time. Introducing the time variable $\tau=\ln (a)$
instead of $t$ (e.g. Barrow \& Silk 1981; Nusser \& Colberg 1998), the
equation of motion reads
\begin{equation} 
\ddot{\vaa}^{\alpha\beta}-(1+q(\tau))\dot{\vaa}^{\alpha\beta}=
-\frac{3}{2}\Omega_m(\tau)\sum_{\gamma}
\Phi^{\alpha\gamma}\vaa^{\gamma\beta}\;,
\label{eq:motion}
\end{equation}
and is manifestly independent of the Hubble constant.  Upper dots
denote derivatives with respect to $\tau$,
$q(\tau)=\Omega_m(\tau)/2-\Omega_\Lambda(\tau)$ is the deceleration
parameter ($q(\tau)=1/2$ in a EdS Universe) and the  gravitational
potential $\Phi$ is in unit of $4\pi G\rb$. $\Omega_m(z)$ and
$\Omega_\Lambda(z)$ are the matter and vacuum density in unit of the
critical density, respectively. The initial  conditions are set by the
Zeldovich approximation (see below). Virialization occurs when the
three axes have collapsed. To prevent axis $k$ from shrinking to
arbitrary small sizes, we halt  its collapse when  $A_k/aR_1=f_r\equiv
0.177$. This freeze-out radius is chosen so that the virialized object
is 178 times denser than the background (Bond \& Myers 1996). When
axis $k$ has collapsed, we set the radial  component of the velocity
and acceleration in that direction to $\dot{A}_k=\ddot{A}_k=f_r a
R_1$. This way we end the radial collapse but leave the tangential
velocity unchanged, so that the angular momentum of the proto\-halo is
conserved (see, e.g., Eisenstein \& Loeb 1995 for details).

\subsubsection{External shear field}

The external force exerted by the rest of the Universe on the triaxial
perturbation may be generically written in terms of the Green function
$G(\vr,\vr')=|\vr-\vr'|^{-1}$. It is natural to adopt spherical
coordinates as perturbations grow from an initially homogeneous and
isotropic background. The potential integral can thus be expanded as a
multipole series (Binney \& Tremaine 1987),
\begin{equation}
\Phi(\vr)=\sum_{lm}\left(2l+1\right)^{-1}\,r^l\, q_{lm}^>(r)\,
Y_l^m(\nvh)\;,
\label{eq:mul1}
\end{equation} 
where $\vr=r\nvh$, $Y_l^m(\nvh)$ are the spherical harmonic functions,
and the coefficients
\begin{equation}
q_{lm}^>(r)=\int \!\!\dd^3r'\,Y_l^m(\nvh')^\dagger\delta(\vr')\,
(r')^{-l-1}
\label{eq:mul2}
\end{equation}
are the multipole moments that characterise the potential at any
(interior) point  $\vr$. The constant and the dipole $l=1$ term,
which corresponds to a translation, do not alter the shape of the
whole region and can be dropped out. The quadrupole $l=2$ term
describes the  force distorting the central region. Analytic
calculation (Quinn \&  Binney 1992) indicate that it dominates the
higher order terms ignored here. This justifies to some extent the
assumption of a quadratic  potential.

The quadrupole $q_{2m}^>$ of the external shear will generally not be
aligned with that of the proto\-halo region (see~\S\ref{sec:ics}).
Although one could  treat the nonlinear evolution of the external mass
distribution with concentric shells (Chandrasekhar 1969; Binney \&
Tremaine 1987; Ryden \& Gunn 1987;  Eisenstein \& Loeb 1995), we adopt
a simpler approach and assume that the proto\-halo is embedded in a
single, large scale triaxial region of initial comoving radius
$R_0>R_1$, not necessarily lined up with the proto\-halo region. In
principle, the proto\-halo will cause the large scale perturbation to
warp, producing in return a non-quadratic potential that breaks the
homogeneity of the small-scale ellipsoid.  To avoid this problem, we
assume that the large scale triaxial perturbation and the background
are unaffected by the collapse of  the proto\-halo and remain
homogeneous at all time (Icke 1973; White \& Silk 1979;  Eisenstein \&
Loeb 1995). We can therefore evolve the large scale  ellipsoid
independently using the model of Bond \& Myers (1996).

\begin{figure}
\center \resizebox{0.45\textwidth}{!}{\includegraphics{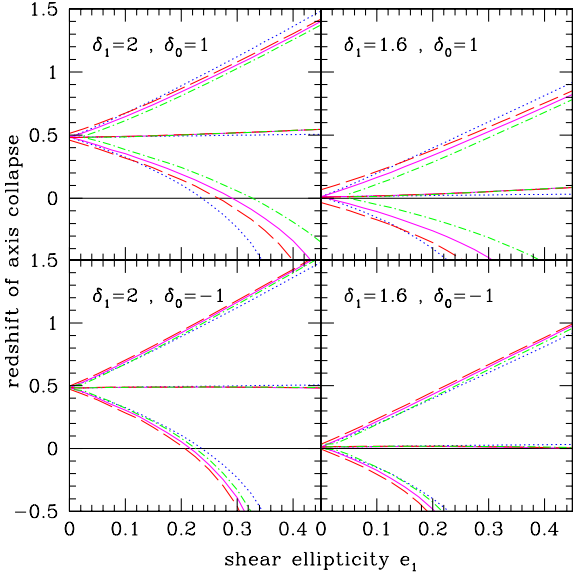}}
\caption{Collapse redshift $z_c$ for the proto\-halo axes as  a
function of the initial ellipticity  $e_1$.  We have assumed a
$\Lambda$CDM Universe.  The  panels show how $z_c$ varies with the
initial proto\-halo overdensity  $\delta_1$ and the large scale
density $\delta_0$ and shear ellipticity $e_0$. In all panels,
$p_0=p_1=0$ and the dotted curve is the reference case
$(e_0,\delta_0)=(0,0)$.  The solid and dashed curves show the collapse
redshifts for $(e_0,\delta_0)=(0,\pm 1)$ and $(e_0,\delta_0)=(0.3,\pm
1)$, respectively. The dotted-dashed curve also has $(e_0,\delta_0)=
(0.3,\pm 1)$, but the initial alignment of the shear is weaker  (see
text).}
\label{fig:zcoll}
\end{figure}

\subsubsection{Gravitational potential of the proto\-halo}

In the principal axis frame of the large scale ellipsoid, we write
down the total gravitational potential of the proto\-halo as
\begin{equation}
\Phi^{\alpha\beta}=\Phi_{\rm FRW}^{\alpha\beta}+ \Phi_{\rm
E,0}^{\alpha\beta}+\Phi_{\rm E,1}^{\alpha\beta}+ \Phi_{\rm
zel}^{\alpha\beta}
\label{eq:tpot}
\end{equation}
where $\Phi_{\rm FRW}^{\alpha\beta}=
\left(1-2\Omega_\Lambda(\tau)/\Omega_m(\tau)\right)/3\,\,
\delta^{\alpha\beta}$ is the contribution of the smooth background and
\begin{eqnarray}
\Phi_{\rm E,0}^{\alpha\beta} \!\!\!\! &=& \!\!\!\!
\frac{1}{2}\delta_0\,b_{0,\alpha}\,\delta^{\alpha\beta} \nonumber \\
\Phi_{\rm E,1}^{\alpha\beta} \!\!\!\! &=& \!\!\!\!
\frac{1}{2}\left(\delta_1-\delta_0\right)\,\sum_{\gamma}
b_{1,\gamma}\,\vqq^{\alpha\gamma}\vqq^{\beta\gamma}
\label{eq:potell}
\end{eqnarray}
are the gravitational potential associated with the large scale
triaxial perturbation and with the remaining mass in the proto\-halo,
respectively. The matrix elements $\vqq^{\alpha\beta}(\tau)$ depend
generally on $\tau$ because the proto\-halo may be rotating.  The
functions $b_\alpha(\tau)$ are defined in Appendix~\S\ref{app:pot},
which provides details on the potential of an homogeneous, triaxial
ellipsoid. It is worth emphasising that, since $b_\alpha(\tau)$ are
homogeneous of degree zero, the potentials~(\ref{eq:potell}) do not
depend on the value of $R_0$ and $R_1$. $\delta_0(\tau)$ and
$\delta_1(\tau)$ are the relative  overdensity of the large scale
environment and the proto\-halo, respectively.  Since the
contribution  of the traceless part of both $\Phi_{\rm
E,0}^{\alpha\beta}$ and $\Phi_{\rm E,1}^{\alpha\beta}$ is of  second
order only, we include the linear approximation for the external shear
field,
\begin{equation} 
\Phi_{\rm zel}^{\alpha\beta}=\sum_\gamma \lambda_{\rm zel,\gamma}
\,\vqq_i^{\alpha\gamma}\vqq_i^{\beta\gamma}\;,
\label{eq:potzell}
\end{equation}
where $\lambda_{\rm zel,\gamma}=D(\tau)\left(\lambda_\gamma-\delta_1/3
\right)$ is a linear function of the initial eigenvalues
$\lambda_\gamma$ of the shear tensor smoothed on scale $R_1$.
$\vqq_i^{\alpha\beta}=\vqq^{\alpha\beta}(\tau_i)$ describes the
initial orientation of the proto\-halo relative to that of the large
scale region. This ensures that the evolution consistently reduces to
the Zeldovich approximation at early times (Bond \& Myers 1996). The
initial conditions are explicitely
\begin{eqnarray}
\vaa^{\alpha\beta}(\tau_i) \!\!\!\! &=& \!\!\!\! a_i R_1
\left(1-D_i\lambda_\beta\right)\,\vqq_i^{\alpha\beta}\nonumber \\
\dot{\vaa}^{\alpha\beta}(\tau_i) \!\!\!\! &=& \!\!\!\!
\vaa^{\alpha\beta}(\tau_i)-a_i R_1 D_i\lambda_\beta\,
\vqq_i^{\alpha\beta}\;,
\label{eq:init}
\end{eqnarray}
where $a_i=a(\tau_i)$ and $D_i=D(\tau_i)$. Note that the initial
tangential velocities are zero.

The simplifications of this model notwithstanding, our calculation
should provide  a quantitatively useful description of the impact of
environment on the   collapse of dense, triaxial regions.

\subsection{Effect of environment on the redshift and critical 
density for collapse}

Our first task is to study the dynamical effect of the large scale
environment on the  collapse of the proto\-halo. Once the cosmological
background is  chosen, the evolution of the proto\-halo is governed by
the initial values of $(e_1,p_1,\delta_1)$, $(e_0,p_0,\delta_0)$ and
the  Euler angles $(\varphi,\vartheta,\psi)$, which describe its
initial orientation relative  to that of the large scale
ellipsoid. For simplification, we choose $p_0=p_1=0$ and set
$(\varphi,\vartheta,\psi)= (0.2,0.2,0.2)$ (in radian). Note that this
particular orientation  has a reasonable probability of occurring
regardless of the configuration shape of the environment (see
Fig.~\ref{fig:pso3}).  The starting redshift is taken to be a
hundred. We store the collapse  redshift $z_c$ of the three axes, as
well as the (linear) critical  density $\dec$ at
virialization. Results are presented for a  $\Lambda$CDM Universe with
$\Omega_m=0.238$ and $\Omega_\Lambda=0.762$.

Fig.~\ref{fig:zcoll} examines how the collapse redshifts $z_c$ of the
three proto\-halo axes change with the model parameters. For reasons
that should become apparent below, $z_c$ is plotted against the shear
ellipticity $e_1$. In all  panels, the dotted curve is the reference
case $(e_0,\delta_0)= (0,0)$. The solid and dashed curves indicate the
collapse redshifts for  $(e_0,\delta_0)=(0,\pm 1)$ and $(e_0,\delta_0)
=(0.3,\pm 1)$,  respectively. To highlight the effect of changing the
relative orientation, we have also plotted as dotted-dashed curves the
collapse redshifts for $(e_0,\delta_0)=(0.3,\pm 1)$ and a weaker
initial shear alignment, $(\varphi,\vartheta,\psi)=(1.5,1.5,1.5)$.
Also shown are the values of $\delta_0$ and $\delta_1$, linearly
extrapolated to present epoch.  For positive values of $\delta_0$, the
tidal force exerted by the large scale perturbation delays the
collapse along the first axis but enhances it along the third,
reducing  thereby the anisotropy that arises from the linear term,
eq.~(\ref{eq:potzell}). By contrast, the difference between the
collapse redshift of the major and minor axis is enhanced for
$\delta_0<0$. In all cases however, the collapse redshift of the
intermediate  axis is barely affected and remains close to the value
predicted by the spherical collapse model (see Appendix of
Shen \etal 2006). Clearly, at fixed
overdensity $\delta_1$, the haloes embedded in the large density
environment virialize earlier.  The strength of this effect increases
with the shear ellipticity $e_1$. It is  very sensitive to the initial
conditions. For $\delta_0=-1$, an external quadrupole shear $e_0>0$
delays the halo virialization as compared to the spherical case. This
delay is still present for $\delta_0=+1$ when the halo collapse is
initially close to spherical (The delay is most significant for the
low value of $\delta_1$), but the trend reverses when the ellipticity
gets larger than   $e_1\bsim 0.1$. In addition, changing the initial
alignment can also increase or decrease the difference in collapse
redshift.  Notice also that, at low redshift,  the increasing
contribution of the cosmological constant to the energy  density slows
down the collapse of the third axis noticeably.

\begin{figure}
\center \resizebox{0.45\textwidth}{!}{\includegraphics{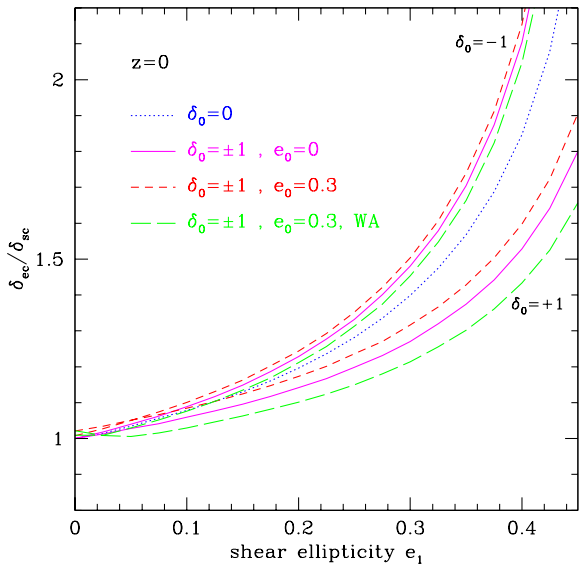}}
\caption{Critical density $\dec(e_1,z)$ for the collapse of a
proto\-halo perturbation at $z=0$ in unit of $\dsc=1.673$, the
critical density  for  a spherical collapse in the $\Lambda$CDM
cosmology considered here.  $\dec$ is plotted as a function of the
shear ellipticity $e_1$. The  various curves indicate how $\dec$
varies for the initial configurations considered in
Fig.~\ref{fig:zcoll}. 'WA' labels the configurations for which the
initial orientation of the shear principal  axis frame is given by
$(\varphi,\vartheta,\psi)=(1.5,1.5,1.5)$ (see text).  The dotted curve
shows  the reference case $(e_0,\delta_0)=(0,0)$. Note  that
$p_0=p_1=0$ in all cases.}
\label{fig:dcoll1}
\end{figure}

When all the parameters but $\delta_1$ are held fixed, then there is a
unique value of $\delta_1=\dec(e_1,z)$~\footnote{In what follows, we
shall omit writing the other variables, but recall that $\dec$ is
generally a function of the 12 variables that parametrise the shear
eigenvalues and the relative orientation of the principal axis frames
on scale $R_0$ and $R_1$.}  which  leads to the collapse  of the minor
axis (virialization) at redshift $z$. In Fig.~\ref{fig:dcoll1}, the
critical density for collapse at $z=0$ is plotted in unit of $\dsc$
for the initial collapse configurations considered in
Fig.~\ref{fig:zcoll}.   We have also set $p_0=p_1=0$. As before, the
dotted curve  shows the  reference case $(e_0,\delta_0)=(0,0)$. The
label 'WA' designates the  configurations that have a weaker initial
alignment, $(\varphi,\vartheta,\psi)=(1.5,1.5,1.5)$.   Clearly, the
critical density $\dec$ is lower for larger  values of
$\delta_0$. This owes to the fact that, at fixed initial  density
$\delta_1$,  haloes that reside  in relatively high density
environments collapse earlier. Furthermore, Fig.~\ref{fig:dcoll1} also
indicates that the environment density $\delta_0$ is the most
influential  parameter. The large scale ellipticity $e_0$, for
instance, has a significant impact on the critical density $\dec$ only
when $\delta_0\bsim 0$. These results suggest that an average critical
density  $B(e_1,z,\delta_0)$ should provide a  good description of
this environmental effect if the scatter around it is not too
large. We investigate this possibility in the remainder of this
Section.

\subsection{Environmental dependence as moving barriers}
\label{sub:moving}

As seen in \S\ref{sec:ics}, the local properties of the shear depend
substantially on the large scale environment.  Here, we consider a
large ensemble of initial halo-environment  configurations and examine
the resultant distribution of critical  density $\dec$.  We find that
it is a reasonable approximation to use an average critical density
$B(e_1,z,\delta_0)$ and neglect the scatter around it. We provide a
fitting formula to $B(e_1,z,\delta_0)$  which facilitates the
inclusion of the environmental dependence of the kind considered here
into the excursion set formalism.

\begin{figure*}
\center \resizebox{0.85\textwidth}{!}{\includegraphics{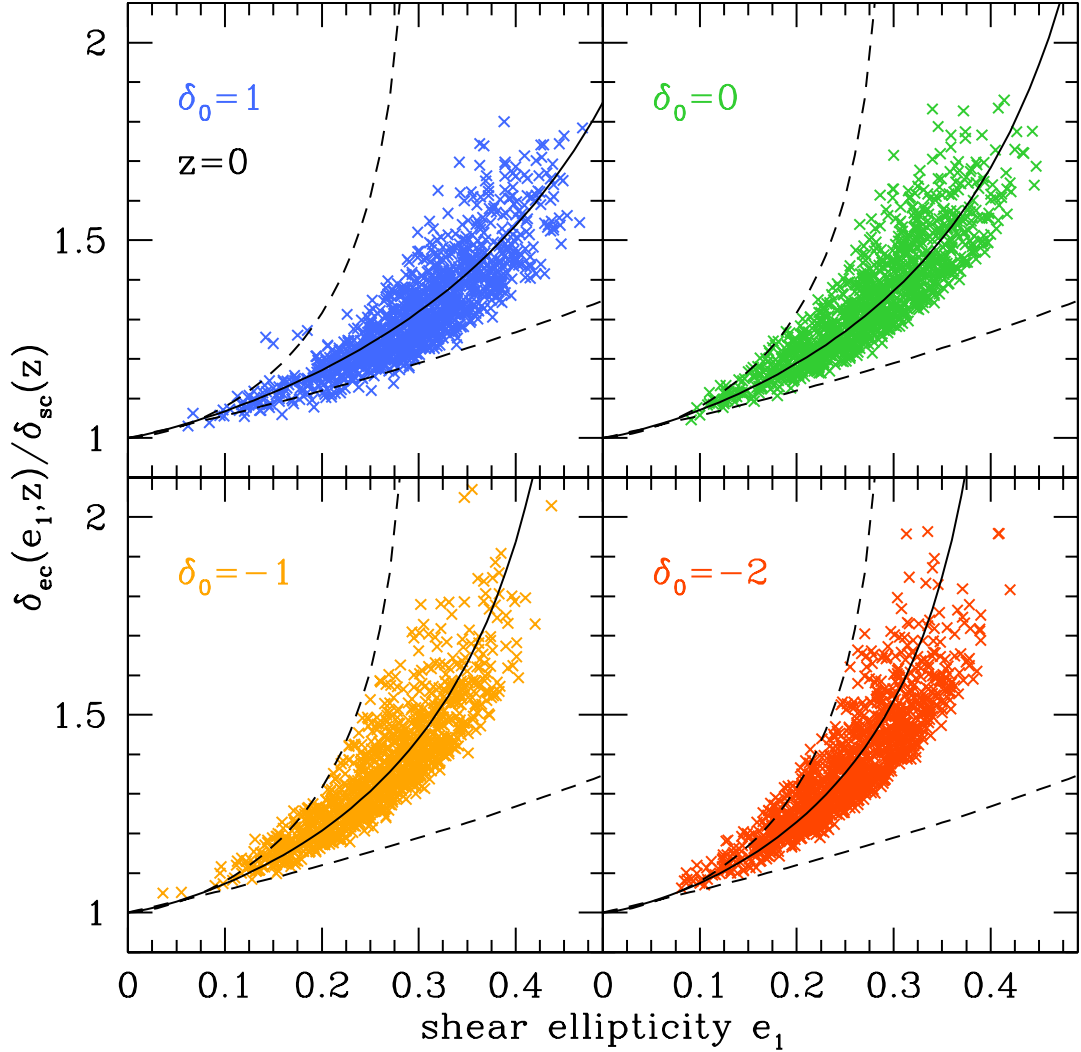}}
\caption{Distribution of critical densities $\dec$ and shear
ellipticities $e_1$ as a function of the environment density
$\delta_0$ for a collapse redshift $z=0$.  The crosses indicate the
actual values of $\dec$ and $e_1$ of $10^4$ individual realisations.
The solid curves show our approximation, eq.~(\ref{eq:barrier}). to
the critical  collapse boundary.  The dashed curves indicate the
(approximate) boundaries of the domain $(e_1,\dec)$ sampled by the
random realisations. They are the barrier shape (\ref{eq:barrier})
with  $\delta_0=\pm 5$.}
\label{fig:dcoll2}
\end{figure*}

\subsubsection{Monte-Carlo simulations}
\label{subsub:mc}

Chiueh \& Lee (2001) have shown that random realisations of the linear
deformation tensor can be simulated by drawing six independent
Gaussian variables. Their algorithm can be easily extended to generate
joint  realisations of the shear tensor  which satisfy the correlation
property~(\ref{eq:corxij}). In the basis defined in
eq.~(\ref{eq:newset}),  the $12\times 12$ covariance matrix $\vmm$
decomposes into a direct sum of $2\times 2$ block-diagonal
matrices. As a result, the variables
$X=\left\{u,v,w,\xi_4,\xi_5,\xi_6\right\}$  can be simulated using the
following linear transformation
\begin{eqnarray}
X_0 \!\!\!\! &=& \!\!\!\!
\frac{\sigma_{_X}}{\sqrt{2}}\left(\sqrt{1+\gamma}\,y_+ -
\sqrt{1-\gamma}\,y_-\right) \nonumber \\ X_1 \!\!\!\! &=& \!\!\!\!
\frac{\sigma_{_X}}{\sqrt{2}}\left(\sqrt{1+\gamma}\,y_+ +
\sqrt{1-\gamma}\,y_-\right) \;,
\label{eq:xigen}
\end{eqnarray} 
where, again, the subscripts 0 and 1 indicate that the shear is
smoothed (with a tophat filter) on scale $R_0$ and $R_1$
respectively. $y_+$ and $y_-$ are two Gaussian random deviate of
dispersion unity. $\sigma_{_X}\equiv \sqrt{\la X^2\ra}$  is the rms
variance in the variable $X$. For instance,  $\sigma_{_X}=1/\sqrt{15}$
when $X=v$. The remaining components of the  shear tensor, $\xi_1$,
$\xi_2$ and $\xi_3$ are readily obtained from the  linear
relations~(\ref{eq:newset}).

Since the small-scale perturbation is identified as a halo at some
redshift $z>0$, while the exterior ellipsoid is assumed uncollapsed at
that redshift, we constrain $\delta_0$ so that it does not exceed
$0.9\,\delta_1$.  We also enforce the constraint $\delta_1\geq
1.6$. For a radius  $R_1=2\hmpc$, this corresponds to a density
threshold $\nu_1\bsim 1$.  In other words, the vast majority of our
haloes form out of one standard deviation fluctuations.  In practice,
we generate random realizations of the initial conditions  and reject
the cases that do not satisfy these constraints.

\subsubsection{Collapse barriers}
\label{subsub:barrier}

We generate a large ensemble of initial collapse configurations.  We
evolve each realisation separately using the model described in
\S\ref{sub:edyn} and store the value of the density contrast
$\delta_1\equiv \dec$ which corresponds to collapse at redshift $z$.

The distribution of critical density $\dec/\dsc$ and shear ellipticity
$e_1$ is plotted in  Fig.~\ref{fig:dcoll2}  for several values of the
primeval environment density, evenly spaced in the range
$-2\leq\delta_0\leq 1$. Collapse occurs at  $z=0$. The crosses
indicate the actual values of  collapse densities and shear
ellipticities of (a small subset of  the) individual
realisations. Also shown as the solid  curve is our approximation to
the critical collapse  boundary  $B\equiv\dec(e_1,z)$ defined by the
implicit equation (Sheth, Mo \& Tormen 2001)
\begin{equation}
\frac{\dec(e_1,z)}{\dsc(z)}=1+\beta_1\left[5\,e_1^2\,
\frac{\dec(e_1,z)^2}{\dsc(z)^2}\right]^{\beta_2}\;.
\label{eq:barrier}
\end{equation}
The functions $\beta_1$ and  $\beta_2$ generally depend on both the
redshift $z$ and the  environment density $\delta_0$. We have adopted
the simple  functional form
\begin{equation}
\beta_i(a,\delta_0)=b_i\,(1+z)^{-d_i}\exp\left(-c_i\delta_0\right)\;.
\label{eq:fform}
\end{equation} 
The exponential factor ensures that $\beta_1$ and $\beta_2$ are
strictly positive. In addition, we have enforced the constraint
$d_i>0$ to account for the (slight) decrease of the environmental
dependence with redshift.   Note also that, in the limit of large
environment densities, the moving barrier $B(e_1,z)$  tends towards
the constant spherical barrier $B=\dsc(z)$. The coefficients $b_i$,
$c_i$ and $d_i$  are found by fitting the barrier shape
(\ref{eq:barrier}) to the critical collapse densities of $3\times
10^5$ realisations with environment density in the range
$-2\leq\delta_0\leq 1$. We find
\begin{eqnarray}
&& b_1\approx 0.412,~~~c_1\approx 0.113,~~~d_1\approx 0.0576 \nonumber
\\ && b_2\approx 0.618,~~~c_2\approx 0.0451,~~~d_2\approx 0.0485 \;.
\label{eq:coeff}
\end{eqnarray}
For $\delta_0=0$ and $a=1$, we obtain $\beta_1=0.412$ and
$\beta_2=0.618$, in good agreement with the values inferred by Sheth,
Mo \& Tormen ($\beta_1=0.47$, $\beta_2=0.615$). A visual inspection of
Fig.~\ref{fig:dcoll2} has convinced us that  the barrier shape
(\ref{eq:barrier}) provides a good approximation to the increase of
the critical collapse density with ellipticity and to its dependence
on the environment density in the range $0\leq e_1\leq 0.45$. To guide
the eye, we have also plotted as dashed curves the (approximate) upper
and lower boundaries of the domain $(e_1,\dec)$ sampled by the random
realisations. These boundaries are the barrier shape
(\ref{eq:barrier}) with an environment  density $\delta_0=\pm 5$.

\begin{figure}
\centering \resizebox{0.35\textwidth}{!}{\includegraphics{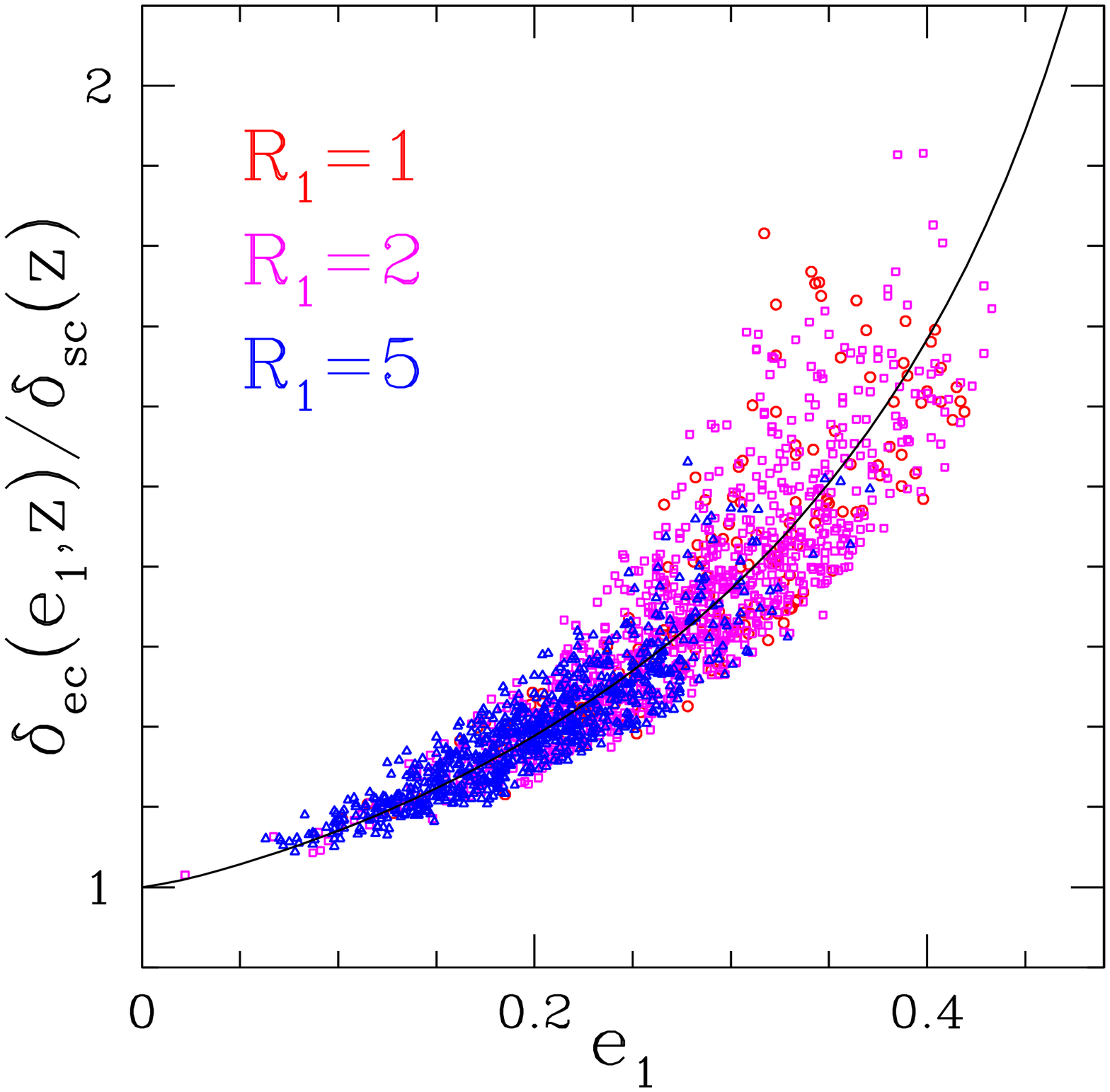}}
\resizebox{0.35\textwidth}{!}{\includegraphics{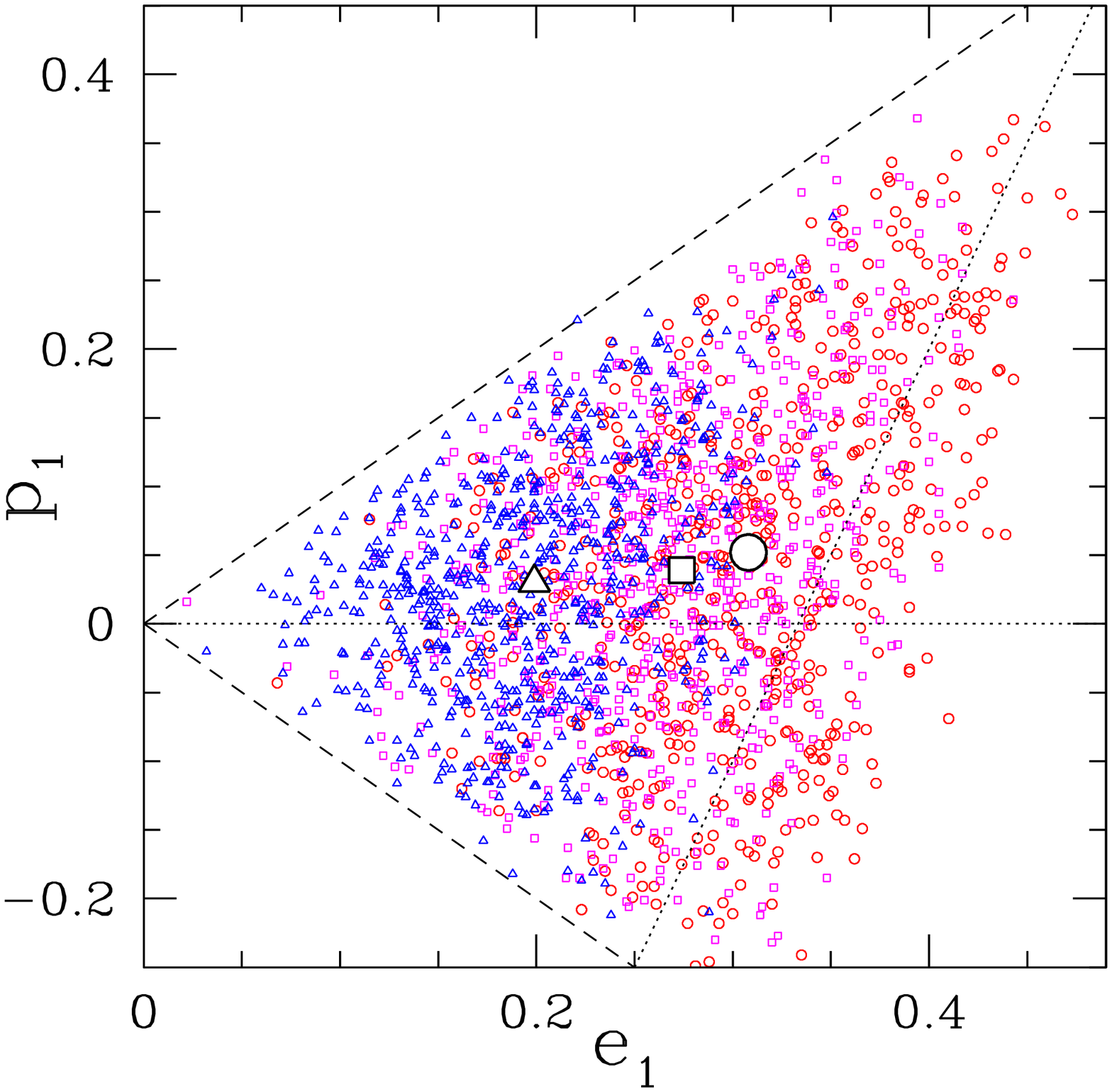}}
\caption{{\it Top panel}~: Distribution of critical density and shear
ellipticity for three different proto\-halo radii $R_1=1$ (circle), 2
(square) and 5$\hmpc$ (triangle). The collapse redshift is $z=0$ and
the environment density is $\delta_0=0$.  The average density for
collapse, eq.~(\ref{eq:barrier}), is plotted as the solid curve. {\it
Bottom panel}~: Distribution of initial shear ellipticity and
prolateness for the individual realizations shown in the top
panel. The interior of the triangle shows the region where
$\lambda_3>0$.  The big empty symbols indicate the peak of the
distributions.}
\label{fig:dcoll3}
\end{figure}

It should be noted that the barrier~(\ref{eq:barrier}) is in good 
agreement with exact result for $p_0=p_1=0$. In this case, the 
parametrization of Sandvik \etal (2007) and Sheth \etal (2001) 
agree fairly well, although the former describes the two-dimensional
surface $\dec(e_1,p_1)$ more accurately. For $\delta_0\neq 0$ however, 
we found that the critical density for collapse depends on the shape 
parameters in a complex way. This is the reason why we resort to the 
Sheth \etal functional form.

Following Sheth \etal (2001), we shall interpret (\ref{eq:barrier}) as
a 'moving' barrier. Such an interpretation has the advantage that,
once the barrier shape is known, the excursion set formalism can be
use to quantify the dependence of halo properties on environment.
Moreover, it is computationally more efficient that studying the first
crossing distribution of multi-dimensional random walks (e.g. Chiueh
\& Lee 2001; Sheth \& Tormen 2002; Sandvik \etal 2007).

\subsubsection{Mass scale-ellipticity relation}
\label{subsub:massscale}

Before we examine the impact of this dynamical interaction on the
properties of collapsed haloes,  we need to express the critical
density (\ref{eq:barrier}) in terms of   the halo mass $M_1$. Thus
far, we have only  considered the collapse of regions with initial
radius $R_1=2\hmpc$.  The top panel of Fig.~\ref{fig:dcoll3} shows the
distribution of  critical densities $\dec$ for three different values
of $R_1$. The average density for collapse is plotted as the solid
curve. Results are shown for an environment density $\delta_0=0$
only. Note, however, that we have repeated this calculation for other
values of $\delta_0$ and found good agreement between the  critical
densities of individual realisations and the approximation
(\ref{eq:barrier}).  This confirms that most of the environmental
effect seen here arises from variation in the large scale density
$\delta_0$. At fixed $R_0$,  changing $R_1$ merely affects the scatter
around the collapse boundary $B(e_1,z)$, unsurprising since the
functions $b_i(\tau)$ that characterise the potential of the
proto\-halo and its environment are independent of $R_0$ and
$R_1$. The decrease in scatter with increasing $R_1$   is a direct
consequence of the statistical correlation explored in
\S\ref{sec:ics}.  This is clearly seen in the bottom panel of
Fig.~\ref{fig:dcoll3}, where the distribution of initial shear
ellipticity and prolateness is plotted as a function of $R_1$.  The
filled symbols indicate the actual, most probable values $(e_m,p_m)$.
They increase (monotonically) with the proto\-halo  radius $R_1$. This
suggests relating the mass scale $M_1$ to the  expectation value of
the asymmetry parameters.

In line with the interpretation of Sheth \etal (2001; see also Shen
\etal 2006; Sandvik \etal 2007), we use the average values
(\ref{eq:em}) to translate the product $e_1\delta_1$ into a peak
height $\nu(M_1,z)=\dsc(z)/\sigma(M_1)$~\footnote{The  peak height
$\nu$ is the typical amplitude of fluctuations that produce haloes of
mass $M_1$ by redshift $z$. A characteristic mass for clustering
$M_\star(z)$ can  then be defined through $\nu(M_1,z)=1$. For the
present cosmology, $M_\star(0)\approx 2.6\times 10^{12}\mdh$.}.  
Gao \& White (2007) have shown that the properties of haloes in the
Millennium simulation obey this scaling relation over  a large
redshift range. In terms of this scaled variable, the collapse
boundary eq.~(\ref{eq:barrier}) becomes
\begin{equation}
B(\nu,z)=\dsc(z)\left[1+\beta_1\left(\nu^2\right)^{-\beta_2}\right]\;.
\label{eq:barrier1}
\end{equation}
Sheth \& Tormen (2002) provide an analytic fit for the first crossing
distribution associated to this barrier shape, which allows us to
easily  calculate halo properties such as formation redshift and bias.

\subsection{Environmental dependence of halo properties}
\label{sub:results}

The main focus is to quantify the environmental  dependence of halo
properties arising from the moving barrier, eq.~(\ref{eq:barrier}).
Even though the ellipsoidal collapse model does not provide an
excellent description of the simulations, it does nevertheless a much
better job than the spherical collapse (Sheth \& Tormen 1999;
2002). Therefore, the  validity of a comparison between predicted and
observed environmental effects should be preserved.  We calculate the
distribution  halo of  halo formation redshift and large scale bias
associated to that collapse boundary. We end this Section with a
discussion of the halo spin parameter.

\subsubsection{Environment density and formation redshift}
\label{subsub:zform}

The halo formation redshift $\zf$ is commonly defined as the epoch at
which the main progenitor has accumulated half of its final
mass. According  to the EPS theory, the probability that the formation
redshift of present-day haloes of mass $M$ (we will henceforth drop
the  subscript 1) is larger than $z$ is given by  (Lacey \& Cole 1993)
\begin{equation}
P(>z_f)=\int_{S}^{S_2}\!\!\dd S'\,\frac{M}{M'(S')}f(S'|S)\;,
\label{eq:czform}
\end{equation}
where $S_2=S(M/2)$. When the excursion set theory is combined to the
ellipsoidal collapse dynamics, the conditional first crossing
distribution $f(S'|S)$ shall be replaced by the following analytic
formula,
\begin{equation}
f(S'|S)=\frac{|T(S'|S)|}{\sqrt{2\pi}(S'-S)^{3/2}}
\exp\left[-\frac{\left(B(S')-B(S)\right)^2}{2(S'-S)}\right]
\label{eq:fs}
\end{equation}
where
\begin{equation}
T(S'|S)=\sum_{n=0}^5 \frac{(S-S')^n}{n!}
\frac{\partial^n\left[B(S')-B(S)\right]}{\partial S^{'n}}\;.
\end{equation}
This Taylor expansion provides a good fit to the conditional
up-crossing  probability for moving barriers of the
form~(\ref{eq:barrier1}) (Sheth  \& Tormen 2002). Note that the
ellipsoidal collapse model predicts too  many haloes with high
formation redshift. The distributions are also  broader than seen in
the simulations (Lin, Jing \& Lin 2003; Giocoli  \etal 2007).

\begin{figure}
\centering \resizebox{0.45\textwidth}{!}{\includegraphics{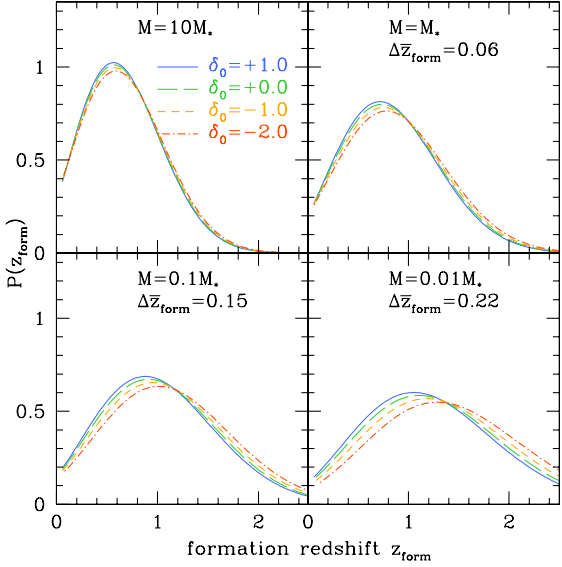}}
\caption{The differential probability distribution of formation
redshift, $P(\zf)$, as a function of halo mass and   (Lagrangian)
environment density $\delta_0$. $P(\zf)$ has been computed from  the
moving barrier~(\ref{eq:barrier1}). The halo mass is given in  unit of
the characteristic mass $M_\star\approx 2.6\times 10^{12}\mdh$.  For
$M\leq M_\star$, we also indicate  $\Delta\mzf$, the difference
between the lowest and highest median  formation redshift.  The mean
formation redshift $\mzf$ increases with  decreasing $\delta_0$.  The
effect is strongest for small mass haloes  $M\ll M_\star$.}
\label{fig:zform1}
\end{figure}

The differential probability distribution $P(\zf)\equiv  \dd
P(>\zf)/\dd\zf$ is plotted in Fig.~\ref{fig:zform1} for various halo
mass $M$ and (Lagrangian) environment density $\delta_0$. Clearly, the
dependence  of halo formation redshift on environment increases with
decreasing halo mass. Equation~(\ref{eq:barrier1}) indeed shows that,
for haloes of mass $M\bsim M_\star$, the critical density for collapse
is $B(\nu,z)\approx\dsc(z)$, weakly dependent on environment density.
This is the reason why the formation redshift of massive haloes is
nearly insensitive to $\delta_0$. By contrast, random walks associated
to small mass haloes $M\ll M_\star$ cross the collapse boundary
$B(\nu,z)$ at relatively larger values of $\nu$ and thus induce a
stronger environment effect. For $M=0.01M_\star$,  we find a median
formation redshift of $\mzf=1.19$ for an environment density
$\delta_0=1$. This should be compared to $\mzf=1.42$ when
$\delta_0=-2$.  Notice that the probability distribution  $P(\zf)$ is
more sensitive to the exponent $\beta_2$ than the multiplicative
factor $\beta_1$. The former contributes about two thirds of the
environmental effect (see~\S\ref{sub:bshape} for more details).

To allow for a direct comparison of our results with the analyses of
N-body simulations, we need to express the environmental dependence of
halo formation redshift in terms of the evolved Eulerian density.  Mo
\& White (1996) and Sheth (1998) have shown how this may be
accomplished within the spherical collapse model. The spherical
collapse dynamics provides a relation between the Lagrangian radius
$R_0$ and density $\delta_0$ and their Eulerian counterparts $R$ and
$\delta$. In this model, the mass interior to each  perturbation is
constant, giving $R_0=R(1+\delta)^{1/3}$ if one assumes that the
primeval density fluctuations are small. Furthermore, the linear
density $\delta_0$ is a monotonically increasing function of the
present overdensity fluctuation $\delta$ only. Mo \& White (1996) have
obtained an accurate approximation to this relation for an EdS
Universe,
\begin{equation} 
\delta_0(\Delta)=\frac{\dsc}{1.686}\left[1.686-\frac{1.35}
{\Delta^{2/3}}-\frac{1.124}{\Delta^{1/2}}+\frac{0.788}
{\Delta^{0.587}}\right]\;,
\label{eq:mw96}
\end{equation}
where $\Delta\equiv 1+\delta$. This interpolation  formula is also
valid in the $\Lambda$CDM cosmology considered here  provided that
$\delta$ is not too large ($\delta\lsim 10$). Ideally, we should
calculate the relative numbers of patches with $(R_0,\delta_0)$ that
have now evolved into regions $(R,\delta)$ (see Sheth 1998). We should
also take into account the triaxiality of the large scale environment
in the conversion of $\delta$ into $\delta_0$. Henceforth however, we
will neglect these complications and use the spherical approximation
to relate the Lagrangian density to the Eulerian density at $z=0$.
This is sensible since, as we have seen, the linear overdensity
$\delta_0$  is the key parameter governing the correlation between
collapse densities and environment. For illustration, the Lagrangian
density is in the  range $-3\lsim\delta_0\lsim 1$ when the  Eulerian
density varies  between $0.3\leq (1+\delta)\leq 5$.

Upon these assumptions, we find that the median formation redshift
changes by
\begin{eqnarray}
&& \Delta\mzf\approx 0.07~~~\mbox{when}~M=M_\star \nonumber \\  &&
\Delta\mzf\approx 0.33~~~\mbox{when}~M=0.01M_\star\;,
\label{eq:zshift}
\end{eqnarray}
when the evolved density varies in the range $-0.7\leq\delta\leq
4$. Again, the statistical correlation induces an effect greater  for
low mass haloes  $M\lsim M_\star$ while, for $M\gg M_\star$, the
offset  in median formation redshift is barely discernible.  In
addition, the difference is of the magnitude seen  in N-body
simulations, where $\Delta\mzf\sim 0.3-0.5$  at $M\sim 0.01M_\star$
for a large scale overdensity varying in  that same range (Harker
\etal 2006). On the other hand, our model predicts that  haloes in
denser regions have a lower formation redshift than those  in less
dense regions. This is opposite to the behaviour reported  by  Harker
\etal (2006), who find that, in overdense regions, haloes  in the
densest environment assemble earliest. A better treatment of  the
relation between Lagrangian and Eulerian regions should not  affect
these conclusions.

Interestingly, however, these authors find that, in the most
underdense regions $\delta\lsim -0.4$, the average  formation redshift
increases with decreasing environment density.  Despite the small
number of haloes and the large scatter in formation redshift, the
trend is robust, yet smaller than the  effects present in the high
density regions (Geraint Harker,  private communication). This may be
a   manifestation of the environmental dependence discussed here.

\subsubsection{The age-dependence of halo bias}
\label{subsub:bias}

Having established a correlation between formation redshift and
environment density, we now turn to the age-dependence of the halo
bias.

\begin{figure}
\centering \resizebox{0.45\textwidth}{!}{\includegraphics{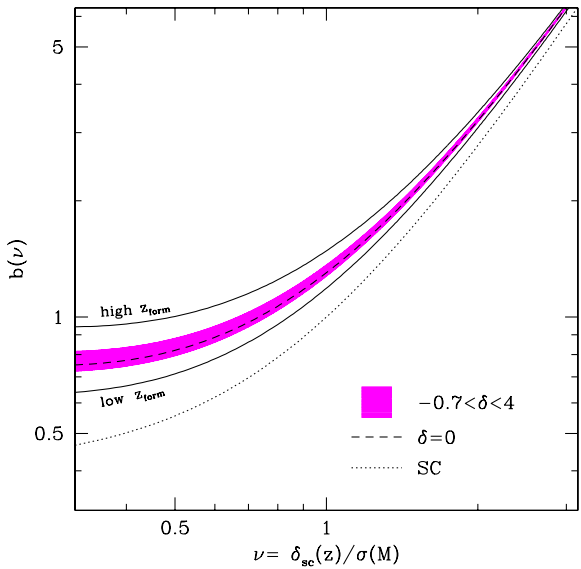}}
\caption{The large scale bias factor $b(\nu)$ as a function of the
peak height $\nu=\dsc/\sigma(M)$. The shaded area indicates the
amplitude of the large scale bias when the Eulerian environment
density varies  in the range $-0.7\leq\delta\leq 4$. The dashed curve
is $b(\nu)$ when $\delta=0$, whereas the dotted line  indicates the
prediction of the spherical collapse. The upper and lower  solid
curves show the bias factor of haloes which have relatively  high and
low formation redshifts (see text for details). Haloes that  assemble
relatively early are more clustered and populate the less  dense
regions. }
\label{fig:bias}
\end{figure}

Mo \& White (1996) and Sheth \& Tormen (1999c) have shown that there
is  a direct relation between the halo bias and the shape of  the
collapse barrier. For a  barrier of the form (\ref{eq:barrier1}), the
large scale bias in  Eulerian space can be approximated by (Sheth
\etal 2001)
\begin{eqnarray}
b(\nu)\!\!\!\! &=& \!\!\!\! 1+\frac{1}{\dsc}\left[\nu^2+\beta_1
\nu^{2-2\beta_2}\right. \nonumber \\ && \left. -\frac{\nu^{2\beta_2}}
{\nu^{2\beta_2}+\beta_1\left(1-\beta_2\right)\left(1-\beta_2/2\right)}
\right]\;.
\label{eq:bias}
\end{eqnarray}
This bias relation is plotted in Fig.~\ref{fig:bias} as a function of
the peak height $\nu$. The dashed curve shows the halo bias at mean
Eulerian density $\delta=0$, namely in the case of the ellipsoidal
collapse of Sheth, Mo \& Tormen (2001), while the dotted line is the
prediction of the spherical collapse. The shaded area indicates the
amplitude of  $b(\nu)$ when the environment density varies in the
range $-0.7\leq\delta\leq 4$. The bias is roughly  10 per cent larger
for the haloes residing in the most underdense  region $\delta=-0.7$.

To compare our results as directly as possible with those of Gao,
Springel \& White  (2005), it would be useful to estimate the bias
factors of haloes which lie in the upper and lower tail of the
formation redshift distribution. In  practice, however, such a
calculation proves difficult in the EPS  formalism. Instead we will
consider a simpler approach motivated by the relation between barrier
height and median formation redshift.  We have so far neglected the
presence of stochasticity in critical collapse densities. In this
respect, Fig.~\ref{fig:dcoll2} shows that, at fixed environment
density, there is a fairly large scatter in $\dec$. To estimate  the
importance of this scatter in the bias, we will also compute $b(\nu)$
for the  barriers that bound the locus traced out by the random
realisations when the environment density varies in  the range
$-0.7\leq\delta\leq 4$ (i.e. a Lagrangian density $-2\leq\delta_0\leq
1$). These upper and lower barriers are shown in Fig.~\ref{fig:dcoll2}
as the dashed  curves. They induce a formation redshift distribution
strongly biased towards the extremes. For $M=0.1M_\star$, the median
formation redshift associated to the upper barrier is larger by
$\Delta\mzf=0.51$ than the value obtained from the lower barrier. We
use these barriers to define our 'old' (high $\zf$) and 'young' (low
$\zf$) haloes. Of course, this is a crude approximation to the 10 (20)
per cent tails considered by Gao \etal (2005), but we have not found
any better alternative.

In Fig.~\ref{fig:bias}, we show the resultant bias relations  as the
solid curves labelled by 'low $\zf$' and 'high $\zf$'.  The relative
bias of our old versus young haloes increases smoothly with decreasing
halo mass. The effect becomes large for $\nu\lsim 1$ because of the
considerably stronger dependence of the barrier shape on the peak
height. In this  regime, the large scale bias of the old haloes is
$\sim$50 per cent larger than for the young ones, an increase roughly
twice as small as seen in the simulations. Overall, the behaviour is
similar to that reported in Gao \etal (2005), where the correlation
between halo clustering and formation history is strong for haloes
less massive than $M_\star$ only.  A  better modelling of the
properties of haloes lying in the tails of the formation redshift
distribution is required to make a more quantitative comparison
between our predictions and the measurements of Gao \etal (2005). The
point of the present analysis is to show that the moving barrier
(\ref{eq:barrier1}) could lead to a correlation similar to that seen
in N-body simulations if the scatter in collapse density is taken
into account.

In contrast to the behaviour seen in N-body simulations, the predicted
bias $b(\nu)$ decreases with increasing environment density,
reflecting the flattening of the barrier shape for large values of
$\delta_0$. In the limit $\delta\gg 1$, $b(\nu)$ tends towards the
value predicted by the spherical collapse. Since,  on average, dense
regions at the present time formed from  relatively dense regions in
the primeval fluctuation field, an  anti-correlation between halo bias
and large scale Eulerian density  is expected in our model. This seems
at first surprising but, as recognised by Abbas \& Sheth (2007), the
large scale bias is not necessarily a monotonically increasing function 
of environment density. Their results strongly suggest that haloes in
extremely underdense environment are more clustered than the mean.

\subsubsection{Alignment of halo spin parameter}
\label{subsub:spin}

The dimensionless spin parameter $\vec{\lambda}$ measures the amount
of rigid rotation acquired by the triaxial perturbation  before
virialization. It is defined as  (Peebles 1969)
\begin{equation}
\vec{\lambda}=\frac{L\sqrt{|E|}}{GM^{5/2}}\vec{\ell}\;
\label{eq:spin}
\end{equation}
where $\vec{L}=L\vec{\ell}$ is the halo angular momentum, $\vec{\ell}$
is a unit vector, and $E=E_{\rm pot}+E_{\rm kin}$, $M$ are the total
energy and mass of the ellipsoid, respectively. The quantities $L$ and
$E$ can be expressed in terms of the matrix elements $A^{\alpha\beta}$
(Peebles 1980; Binney \& Tremaine 1987; Eisenstein \& Loeb 1995). In
general, the energy $E$ of the collapsing region is not conserved,
notably  because the kinetic energy of the ellipsoid is altered when
an  axis collapses. For simplicity, we use the last value of the
energy before the first axis collapses in the calculation of
$\lambda=|\vec{\lambda}|$ (Eisenstein \& Loeb 2005). This is a good
approximation as the change in total  energy is usually small during
the collapse. Note also that the resulting error should be relatively
small because $\lambda$ depends on $\sqrt{|E|}$ solely.

\begin{figure}
\centering \resizebox{0.45\textwidth}{!}{\includegraphics{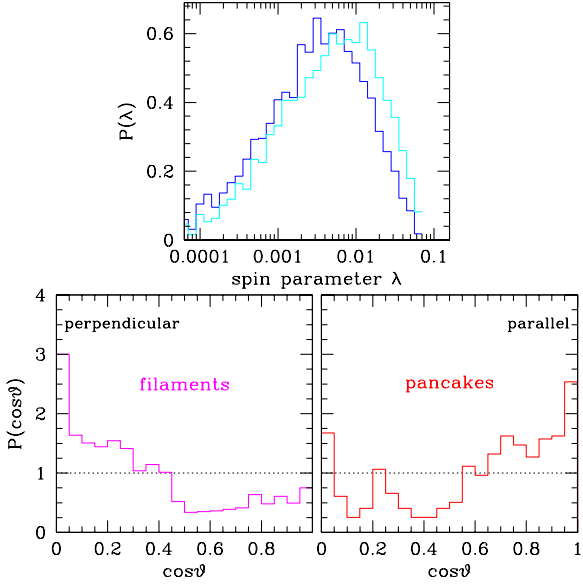}}
\caption{{\it Top panel}~: Effect of including the correlation between
the initial orientation of the proto\-halo and large scale environment
on the probability distribution  of the spin parameter $\lambda$.  The
left histogram shows $P(\lambda)$ when the correlations in the initial
alignment are included, the right histogram when they are not.   {\it
Bottom panels}~: Probability distribution of the cosine of the angle
between the angular momentum of the collapsed halo and both filament
direction (bottom left) and sheet normal vector (bottom right). A
random  distribution would be a flat line at
$P(\cos\vartheta)=1$. Results are shown for haloes that collapse at
$z=0$. The histograms were drawn from $\sim 10^4$ random realizations
of the initial conditions. }
\label{fig:spin}
\end{figure}

The top panel of Fig.~\ref{fig:spin} illustrates the effect of
including  the correlation between the initial orientation of the
proto\-halo and its large scale environment on the probability
distribution $P(\lambda)$  of the spin parameter. The left histogram
shows $P(\lambda)$ when  the correlation in the initial alignment is
included, the right  histogram when statistical independence is
assumed. The distributions were drawn from $10^4$ Monte-Carlo
realizations of the initial  conditions. Clearly, the inclusion of
correlations in the alignment  of the principal axes has a strong
impact on $P(\lambda)$~: It lowers the median  spin by $\lsim 40$ per
cent. As pointed out by several authors (Steinmetz \& Bartelmann 1994;
Catelan \& Theuns 1996; see also Hoffman 1988), the assumption of
random relative  orientations overestimates  the growth of angular
momentum because, if the principal axis frames  tend to be partially
aligned, the angular momentum gain is reduced.  Fig.~\ref{fig:spin}
confirms that the correlation in the primeval  alignment is an
important factor, in agreement with the analysis of Lee \& Pen  (2001).

Our median spin value is $\lambda_{\rm med}\approx 0.005$, an order
of magnitude lower than those found in numerical simulations, where
$\lambda_{\rm med}\sim 0.03-0.05$ for haloes of mass  $M\bsim
10^{11}\mdh$ (Barnes \& Efstathiou 1987;  Bullock \etal 2001; Bett
\etal 2007). There are several reasons for  this discrepancy. First,
the number density of haloes of a given  mass depends noticeably on
the environment density $\delta$~: haloes  are preferentially found in
mildly overdense regions. Here, however,  the distribution of evolved
density $\delta$ associated to the  random realisations of
Fig.~\ref{fig:spin} is not representative  of a fair  halo sample~: it
peaks around $\delta=0$, where the average spin at collapse is much
lower than in high density regions $\delta\bsim 1$. Second, the
angular momentum of a collapsing region satisfies the equation
\begin{equation}
\dot{L}_i=-H^{-1}\epsi_{ijk}{\rm T}^{kl}{\rm I}^{lj}\;,
\label{eq:ang}
\end{equation}
where ${\rm T}^{ij}=(3/2)\,H^2\Omega_m (\Phi_{\rm E,0}^{ij}+\Phi_{\rm
zel}^{ij})$ is the torque and ${\rm I}^{ij}=(1/5)M
(\vaa\vaa^\top)^{ij}$ is the inertia tensor.  We note that the angular
momentum  vanishes to first order if the inertia tensor is zero at
initial time, or if ${\rm I}$ and ${\rm T}$ are perfectly aligned
(White 1984; Catelan \& Theuns 1996). In this limit, the growth of
angular  momentum is of second order only, $L\propto a^{5/2}$, and is
dominated by nonlinear effects after turnaround (Peebles 1969).  This
is indeed the case here as we are considering perturbations  that are
initially spherical.  A more realistic treatment should include
first-order tidal torquing. Third, our axis collapse condition
strongly reduces the amount of angular momentum gained by the halo
after the second axis collapses. A different  prescription, such as
the one adopted by Eisenstein \& Loeb (1995),  would increase the
magnitude of the spin parameter at third axis collapse.  Finally,
analytic calculations indicate that a significant fraction of the
angular momentum is acquired through the collapse of the outermost
shells (Ryden 1988; Quinn \& Binney 1992). This can only be taken into
account by a detailed modelling of the matter density and velocity
profiles  around the collapsing haloes.

In spite of these limitations, it is worthwhile looking at the
alignment  between halo spin and environment as this quantity does not
depend on the magnitude of the angular momentum gained during the
collapse.  In the bottom panels of Fig.~\ref{fig:spin}, the histogram
shows  the alignment distribution  at collapse time. The angle is
measured between the angular momentum of   the collapsing region and
the symmetry axis/plane of the mass  distribution.   Clearly, haloes
show a strong tendency to have their spin aligned  perpendicularly to
the filament or parallel to the mass sheet. This is in good agreement
with the findings of Hahn \etal (2007) and Arag\'on-Calvo \etal 
(2007a), although the former authors found 
a clear correlation for haloes residing in sheet-like  structures only 
and the latter a mass-dependent spin orientation in filaments
(see also Sousbie \etal 2007; Trujillo, Carretero \& Patiri 2006). 
More precisely, the alignment is
strongest along the second principal axis of the shear tensor. This
owes to the fact that, once the first axis has collapsed,
$\dot{L}_i\sim \epsi_{ijk}\,\Phi_{\rm E,0}^{kl}\,{\rm I}^{lj}$.  The
growth rate is largest for the intermediate axis, $\dot{L}_2\propto
(\alpha_1-\alpha_3)\,{\rm I}^{23}$, since the difference
$\alpha_1-\alpha_3$~\footnote{Here the $\alpha_i$s are  the (ordered)
eigenvalues of the large scale potential $\Phi_{\rm E,0}^{ij}$}
dominates the other two (see also Lee \& Pen 2000).  It is unclear
whether the initial alignment between inertia and deformation tensors
reported by Lee \& Pen (2000) and Porciani, Dekel \& Hoffman   (2002b)
can produce a similar correlation. We have not investigated  this issue
any further.

\vspace{6pt}

To summarise, we have demonstrated that our simplified model, which
takes into account both the dynamical and statistical aspects of the
ellipsoidal collapse, produces a clear correlation between formation
redshift, large scale bias and environment density. The strength of
the effect is of the same magnitude as seen in simulations.  It is
largest for low mass haloes, $M\ll M_\star$, and fades as we go to
high masses, $M>M_\star$. Haloes that formed at high redshift are
substantially more clustered than those that assembled recently. This
is precisely the behaviour reported by Gao, Springel \& White
(2005). On the other hand, this model predicts a negative correlation
between formation redshift and environment density, in contradiction
with the trend measured by Harker \etal (2006).  However, simulations
indicate that halo properties depend on environment in a complex way~:
in relatively underdense regions, the average formation redshift
increases with decreasing environment density (Harker \etal 2006)
while haloes are more strongly clustered than the mean (Abbas \& Sheth
2007).  Our model produces an effect in the right sense. This suggests
that the ellipsoidal collapse may apply in underdense regions where
nonlinear effects are weak or absent.

\section{Discussion}
\label{sec:discussion}

In this Section, we discuss the sensitivity of the environmental
effects considered above to the shape of the collapse barrier, as well
as non-Markovianity and tidal interactions as potential sources of
environmental dependence.

\subsection{Sensitivity to the barrier shape}
\label{sub:bshape}

\begin{figure}
\centering \resizebox{0.45\textwidth}{!}{\includegraphics{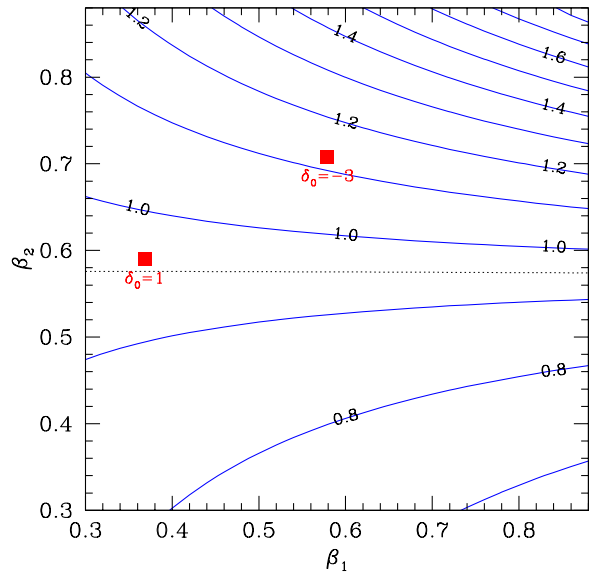}}
\caption{Sensitivity of the median formation redshift to the shape
parameters $\beta_1$ and $\beta_2$. Contours of constant $\mzf$ are
plotted for a halo mass $M=0.1M_\star$. The contour levels are  evenly
spaced, with an interval $\Delta\mzf=0.1$. The filled symbols indicate
the median formation redshifts for the two extreme cases $\delta_0=1$
and $-3$ when $\beta_1$ and $\beta_2$ assume the functional
form~(\ref{eq:fform}). The effect of varying $\beta_1$ vanishes along
a critical line $\beta_2=\beta_2^c$, where  $\beta_2^c\simeq 0.58$
(dotted curve).}
\label{fig:cslope}
\end{figure}

The strength of the environmental effect explored in
\S\ref{sub:results}  may somewhat depend on the  parametrisation
adopted for the moving barrier. While this is probably  true for
haloes of mass $M\bsim M_\star$, the detailed shape of the   barrier
should have a little impact when $M\lsim M_\star$. In this mass range,
the average collapse density $B(\nu,z)$, a Taylor series without  loss
of generality, should be dominated by the term of largest degree.  For
the simple case considered here,  $B(\nu,z)\propto \beta_1
\nu^{-2\beta_2}$ when $\nu\gg 1$.
 
The exponent $\beta_2$ has a great influence on the  correlation
between formation redshift and environment density.  This is clearly
seen in Fig.~\ref{fig:cslope}, where contours of  constant $\mzf$ are
plotted for a halo mass $M=0.1M_\star$.  Shown for illustration are
the median formation redshifts of two  extreme cases $\delta_0=1$ and
$-3$ (filled symbol) when $\beta_1$  and $\beta_2$ assume the
functional form~(\ref{eq:fform}). The  parameter $\beta_1$ has a
substantial impact on the formation redshift only when $\beta_2$  is
far from a critical (empirically determined)  value $\beta_2^c\simeq
0.58$. Along  this curve, the effect of varying $\beta_1$ vanishes,
presumably  because the effective spectral index $n_{\rm eff}$ that
controls  the shape of $M(S)$ in eq.~(\ref{eq:czform}) conspires to
maintain  the median formation redshift constant. Since $n_{\rm eff}$
varies with the mass scale, we expect $\beta_2^c$ to change somewhat
with  the halo mass. The fact that the ellipsoidal dynamics predicts a
value $\beta_2=0.618$ close to $\beta_2^c$ is a coincidence.

Regarding the environmental dependence of halo bias, note that, in the
limits $\nu\gg 1$ and $\nu\ll 1$, the bias offset is $\Delta b(\nu)
\approx -2\beta_1\Delta\beta_2\,\ln\nu\,\nu^{2-2\beta_2}/\dsc$ to
first  order in $\Delta\beta_1$ and $\Delta\beta_2$.  In this regime,
changing the value of $\beta_2$ has a large  effect on the bias of
haloes lying at the extreme ends of the mass range. However, unlike
the environmental dependence of formation redshift,  it is the
parameter $\beta_1$ that influences most the bias when the halo mass
is in the range $0.1\lsim M/M_\star\lsim 10$. We also note that the
fitting formula~(\ref{eq:bias}) of Sheth, Mo \& Tormen (2001) holds
for $\beta_2$ strictly less than one only. It would be prudent to
check again their approximation, namely, derive the first crossing
distribution from a large ensemble of random trajectories when the
values of $\beta_1$,  $\beta_2$ differ significantly from those at
mean density ($\delta_0=0$).

We have shown that each of the shape parameters $\beta_1$ and
$\beta_2$ has a distinct impact on the halo formation redshift and
large scale bias. Therefore, an environmental dependence  of formation
redshift and bias will be present regardless the exact values of these
variables.

\subsection{Environmental effect from non-Markovianity}
\label{sub:corrwalks}

In the excursion set formalism,  a spherical symmetric window function
$\hw(R,k)$ is used to define the trajectories of the linear density
field  $\delta(R)$ as a function of smoothing scale. When the sharp
$k$-space filter is adopted, $\delta(R)$ executes a random walk. The
property of Markovianity has been exploited extensively to obtain
analytic expressions for the first crossing distribution etc. This is
the reason why we have considered that particular window function in
the computation of halo formation redshift and bias.  However, the
Markov nature of Brownian walks prevents any correlation  between
large scale environment and assembly history (see White 1996  for a
discussion). Halo merger trees are indeed non-Markovian across short
time steps (Neistein \& Dekel 2007).
A natural way to introduce correlations would be to use
another window function. This possibility has been considered by
several authors (e.g. Peacock \& Heavens 1990; Bond \etal 1991;
White 1996; Sch\"ucker  \etal 2001; Nagashima 2001; 
Amosov \& Sch\"ucker 2004; Zentner 2007).
Here we argue that non-Markovianess should lead to an effect that is
larger for high mass haloes.

For concreteness, let us consider the collapse of a perturbation of
comoving size $R_1$. To investigate the impact of environment on its
formation history, we restrict  ourselves to trajectories that obey
the following two  constraints~:
\begin{eqnarray}
C_0:~ \delta(R_0) \!\!\!\!\! &=& \!\!\!\!\! \nu_0\,\sigma_0 \nonumber
\\ C_1:~ \delta(R_1) \!\!\!\!\! &=& \!\!\!\!\! \nu_1\,\sigma_1 \;,
\label{eq:c1}
\end{eqnarray}
where, again, $\sigma_i=\sigma(R_i)$ and $R_0=10\hmpc$ is the scale
of the large scale environment. We neglect triaxiality and choose
$\nu_1=\dsc/\sigma_1$ to ensure that the halo has  just collapsed by
redshift $z=0$. $\nu_0$ can be positive or negative, depending on
whether  the halo forms in a high or low density region.

We can calculate the most probable trajectory $\bar{\delta}(R)$ given
the constraints $\left\{C_i\right\}$. Since these are linear
functional  of the density field, the probability of possible
realization  $\delta(R)$ can be expressed as a shifted Gaussian around
an ensemble  mean field (see Adler 1981; Hoffman \& Ribak 1991; Van de
Weygaert  \& Bertschinger 1996 for a rigorous treatment)
\begin{equation}
\bar{\delta}(R)=\zeta_i(R)\,\zeta^{-1}_{ij}c_j\;,
\label{eq:meanf}
\end{equation}
where $\zeta_i(R)=\langle \delta(R)\,C_i\rangle$ is the
cross-correlation between  the field and the $i$th constraint, and
$\zeta_{ij}=\langle C_iC_j\rangle$  is the constraints' correlation
matrix. The residual field  $\tilde{\delta}=\delta-\bar{\delta}$ is a
Gaussian random field which  is not homogeneous nor isotropic, but
whose statistical properties are  independent of the
$\left\{C_i\right\}$ (Hoffman \& Ribak 1991). We define a normalised
cross-correlation between the constraints and the field,
\begin{equation}
\zeta_i(R)\equiv \frac{1}{\sigma_i\sigma} \int_0^\infty\!\!\dd\ln
k\,\dkk \hw(R_i,k)\hw(R,k)\;,
\end{equation}
where $\sigma\equiv\sigma(R)$ and $\hw(R,k)$ is the tophat filter. The
calculation of the matrix elements $\zeta_{ij}$ is immediate. The mean
field $\bar{\delta}$ can be expressed as
\begin{equation}
\bar{\delta}(R)=  \frac{\nu_0\,\sigma}{\left(1-\gamma^2\right)}
\left\{\zeta_1\left(\frac{\nu_1}{\nu_0}-\gamma\right)
+\zeta_0\left(1-\gamma\,\frac{\nu_1}{\nu_0}\right)\right\}\;.
\label{eq:meand}
\end{equation}
The variance of the residual  is
$\tilde{\sigma}^2(R)=\sigma^2-\zeta_i\,\zeta^{-1}_{ij} \zeta_j$ and
does not depend on the constraints $C_0$ and $C_1$. For a  given mass
scale $M_1\propto R_1^3$, the  difference $\bar{\delta}_+-
\bar{\delta}_-$ between the mean field of two different large scale
environments with density contrast $\nu_+$ and $\nu_-<\nu_+$ is
\begin{equation}
\bar{\delta}_+(R)-\bar{\delta}_-(R)=\left(\nu_+-\nu_-\right)\,\sigma\,
\frac{\left(\zeta_0-\gamma\,\zeta_1\right)}{\left(1-\gamma^2\right)}\;.
\label{eq:diffd}
\end{equation}
This difference is negative  (positive) when the smoothing radius is
$R<R_1$ ($R>R_1$). In other words, the density profile around
$\delta(R_1)$ is steeper in low  density regions. Note that, for a
sharp $k$-space filter,  $\bar{\delta}_+-\bar{\delta}_-$ vanishes when
$R<R_1$ since  $\zeta_1=\zeta_0/\gamma$, but is generally non-zero in
the range  $R>R_1$.

\begin{figure}
\centering \resizebox{0.45\textwidth}{!}{\includegraphics{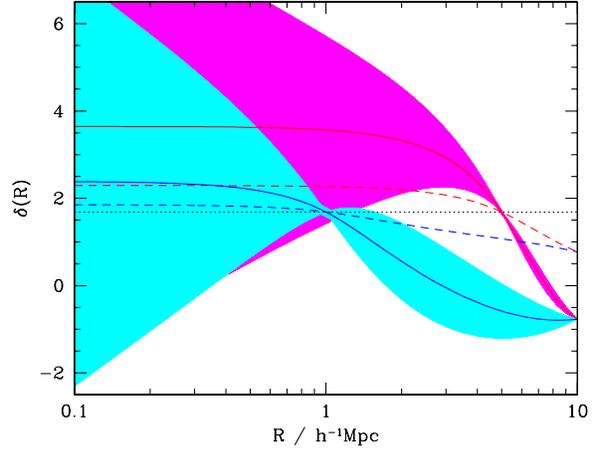}}
\caption{Mean field $\bar{\delta}(R)$ as a function of the smoothing
radius $R$ for two different halo size $R_1=1$ and $5\hmpc$.  The mean
field obeys the following two  constraints~: i) $\delta(R_1)=\dsc$ ii)
$\delta(R_0)=\pm\sigma_0$. The  solid and dashed curves show the
average trajectories which satisfy  $\delta(R_0)=-\sigma_0$ and
$\delta(R_0)=+\sigma_0$, respectively.  The shaded areas indicate the
68\% scatter around the mean. For clarity, they are  attached  to the
curves with $\delta(R_0)=-\sigma_0$ only. The horizontal line  is the
constant spherical collapse barrier $\delta=\dsc$. Results are  shown
for the tophat filter.}
\label{fig:walk}
\end{figure}

Fig.~\ref{fig:walk} shows the mean field $\bar{\delta}(R)$ for the
tophat filter. $\bar{\delta}(R)$ is plotted as a function of the
smoothing radius for two different halo mass $M_1=2.7\times 10^{11}$
and $3.4\times 10^{13}\mdh$. On larger scale $R_0=10\hmpc$, the
density  contrast is set to $\nu_+=1$ (dashed curves) and $\nu_-=-1$
(solid  curves).  To ensure that the trajectories $\delta(R)$ describe
the formation  history of $M=M_1$ haloes, we should also have
constrained the density  so that $\delta(R)$ does not cross the
barrier $\delta=\dsc$ on scale larger than $R_1$.  Here we ignore this
constraint since its implementation is not  straightforward  when the
window function differs from a sharp $k$-space filter. This is the
reason why a substantial fraction of the trajectories penetrate the
barrier $\delta=\dsc$ on scale $R>R_1$. This caveat notwithstanding,
the present calculation is adequate to understand, at least
qualitatively, the effect of non-Markovianity.  Notice that the
difference $\bar{\delta}_+ -\bar{\delta}_-$ is  approximately constant
throughout most of the halo formation history  ($R/R_1\leq 1$). Most
importantly, the effect is about twice as large  for the high mass
halo while, at fixed $R/R_1$, the scatter of the  residual field is
lower by $\sim 60$ per cent.

This suggests that the environmental dependence which arises from
non-Markovianess is stronger for high mass haloes. This has also been
pointed out by Zentner (2007). Therefore,  it is unlikely to explain
the trend seen in overdense regions where,  undoubtedly, correlations
are stronger for low mass haloes (e.g. Gao \&  White 2007). However,
it may apply for isolated haloes in voids or  underdense regions. It
should also leave a signature distinct from the ellipsoidal collapse
which,  as we have seen, induces a stronger dependence for low mass
haloes.

\subsection{On tidal interactions}
\label{sub:tidal}

Irregularities in the mass distribution induce non-radial motions that
slow down the collapse (Peebles \& Growth 1976; Davis \& Peebles 1977;
Peebles 1990). This ``previrialisation'' conjecture is supported by
the numerical investigations of, e.g., Barrow \& Silk (1981); Szalay
\& Silk (1983); Villumsen \& Davis (1986); Lokas \etal (1996); and by
the analytic calculations of Del Popolo \etal (1998; 2001), which
indicate that tidal heating can counterbalance the effect of the shear
and delay the collapse.  Recently, Avila-Reese \etal (2005),
Maulbetsch \etal (2006), Wang \etal (2006) and Diemand \etal (2007)
have proposed that the assembly bias seen in N-body simulations
originates from tidal interactions with a larger neighbour. They have
shown that, at late time, the tidal field of massive neighbouring
clumps halts the growth of haloes and, in many cases, even reduces
their mass.  Furthermore, tidal effects appears to have a larger
influence on small mass haloes. Therefore, this could also explain why
the age-dependence of clustering is stronger for low mass haloes. The
large effect measured by Diemand \etal (2007) indicates that, in high
density regions, tidal interactions are likely to overwhelm the
environmental dependence arising from anisotropic collapse, and
increase the average formation redshift in overdense regions.

Although the impact of tidal stripping can only be rigorously
quantified with numerical simulations, the suppression of mass
accretion through tidal heating could also be addressed analytically.
The spherical collapse model, in which the collapsing object is
divided into a series of concentric shells, seems better suited than
the ellipsoidal collapse. A thorough discussion of tidal  heating is
beyond the scope of the present paper.  Note, however, that the torque
imparted by the (external) mass  distribution on a thin spherical
shell is ${\bf\tau}
(\vx)\propto\int\dd\Omega\,\tilde{\delta}(\vx)\wedge\grad\phi(\vx)$
(e.g. Ryden \& Gunn 1987), where
$\tilde{\delta}(\vx)=\delta(\vx)-\bar{\delta}(x)$ is the deviation
from the spherically symmetric distribution $\bar{\delta}(x)$. This is
$\tilde{\delta}$ which  pulls the infalling matter out of its  purely
radial motion. In the linear regime, $\tilde{\delta}$ is a  Gaussian
density field statistically independent of the spherical average
$\bar{\delta}(x)$. Hence, tidal heating does not induce any
environmental dependence at first order. Consequently, modelling the
growth of nonlinearities in the surrounding mass  distribution will be
crucial  to ascertain analytically the importance of tidal heating in
the environmental dependence of halo collapse.

\section{Conclusion}
\label{sec:conclusion}

The ellipsoidal collapse model is an extension of the spherical
dynamics that takes into account the anisotropic collapse of triaxial
perturbations.  This  non-spherical dynamics provides a substantially
better description of  halo statistics such as mass function and large
scale bias. It is, however, unclear whether the ellipsoidal collapse
can induce environmental effects similar to that seen in N-body
simulations.

In this paper, we have attempted to address this issue, paying special
attention to both the statistical and dynamical origin of the
environmental dependence. In a first part, we have explored the
statistical correlation that arises in (Gaussian) initial conditions
between the local properties of the shear and the configuration of the
large scale environment. To this purpose, a number of joint statistics
for the shear tensor have  been derived, thereby extending the
previous analysis of Doroshkevich  (1970). In a second part, we have
examined the dynamical aspect of the  environmental  dependence using
a simplified model that takes into account the interaction between a
collapsing, ellipsoidal perturbation and its large scale environment.
Relaxing the assumption of sphericity (at the heart of the spherical
collapse) introduces a dependence of collapse redshift on
environment. The tidal force exerted by the surrounding mass
distribution alters the collapse of the major and minor axes, and
causes haloes embedded in large overdensities to virialize
earliest. We have found that the  environment density is a key
parameter in determining the virialization redshift, the large scale
asphericity contributing mostly to increase the scatter in collapse
density. An effective barrier whose shape depends on the large scale
density provides a good description of this environmental effect.
Such an interpretation has the advantage that the EPS formalism can be
applied to estimate the environmental dependence of halo properties
like formation redshift and large scale bias.

We have shown that, using this moving barrier approach, a correlation
between  formation redshift, large scale bias and environment density
naturally  arises. The magnitude of the effect is similar, albeit
smaller, to that seen in N-body simulations.  It is large  for low
mass haloes $M\ll M_\star$, and fades as we go  to high masses
$M>M_\star$ as a result of a genuine statistical effect, namely,  the
decrease in average asymmetry and  stochasticity with increasing halo
mass. Haloes that formed at high redshift are found to be more
clustered than those that  assembled recently. This is precisely the
behaviour reported by Sheth \& Tormen (2004); Gao, Springel \& White
(2005). However, haloes in denser regions are predicted to assemble
later. This result is inconsistent with the trend measured in
overdense environments $\delta\bsim 0$. It calls into question the
role of the ellipsoidal collapse in shaping the halo mass function.
Nevertheless, several lines of recent evidence indicate that, in
relatively underdense regions,  the average formation redshift
increases with decreasing environment  density (e.g. Harker \etal
2006), while haloes may be more strongly clustered than the mean
(Abbas \& Sheth 2007). Our model predicts and effect in the right
sense.  This suggests that the ellipsoidal collapse model may be
applicable in underdense regions, where tidal interactions are weak or
absent.  Conditional halo mass functions $n(M|\delta)$ could provide
another testable prediction since, in the moving barrier
interpretation adopted  here, it is expected that $n(M|\delta)$ in
underdense regions should be (slightly) biased  towards high mass
haloes as compared to the prediction of Sheth \&  Tormen (2002).

Recently, Sandvik \etal (2007) have discussed a multi-dimensional
extension of the EPS formalism that takes into account ellipsoidal
collapse (Chiueh \& Lee 2001).  They find a very weak correlation
between halo assembly history and environment, presumably because
their implementation includes only the statistical aspect of
environmental effects. In our scenario, the dynamical interaction
between the external mass distribution and the collapsing halo plays a
crucial role in the environmental dependence of halo properties. The
statistical correlations contribute mostly to increase the strength of
the effect with decreasing halo mass.  Keselman \& Nusser (2007) have
shown that the age-dependence of halo bias persists in a simplified
description of gravitational dynamics, suggesting thereby that (at
least some of) the environmental dependence arises in the early stages
of the collapse. They find that halo formation redshift strongly
correlates with the dimensionless parameter $\eta\propto (1+6 e_1^2+2
p_1^2)^{-1/2}$, young haloes forming from fluctuations with  higher
$\eta$ than old ones. They argue that this follows from the dependence
of {\it first axis} collapse on configuration shape, namely, a planar
perturbation collapses faster than a spherical perturbation with the
same initial density (Bertschinger \& Jain 1994). Their results are,
however, difficult to reconcile with the above interpretation since
it implies that, for a fixed halo mass, haloes forming in regions of
larger $\eta$ are older (because they grow less rapidly). This
discrepancy can be alleviated if one assumes that the formation of
virialized haloes corresponds to the collapse of the {\it third
axis}. In this case, the collapse is always faster for spherical
perturbations (Audit, Teyssier \& Alimi 1997) so that young haloes
tend to form in regions of higher $\eta$, i.e. more spherical or
denser environments, in agreement with our findings.

A serious shortcoming of  our model is the neglect of anisotropies
beyond the quadrupole term in the external density field. Furthermore,
apart from  a global rotation, it ignores the non-radial degrees of
freedom  in the collapsing proto\-halo. It would be of great interest
to ascertain whether non-radial  motions created by a clumpy, growing
large scale mass distribution can induce an effect similar to that
seen  in simulations. {\it A priori}, tidal heating should be more
efficient in high density environments. It may plausibly reverse the
trend found in this paper, namely, increase the critical collapse
density in large scale overdensities. If this happens to be true, a
moving barrier approach would naturally predict haloes in dense
regions to assemble relatively early and to be more strongly clustered
than the mean. Alternatively, Wang \etal (2007) have suggested that
tidal heating causes haloes to appear less massive than expected from
their initial density field. This may also provide an explanation for
the relatively large/low bias of old/young haloes.  Clearly, large
N-body simulations are needed to understand the complex relations
reported by, e.g. Harker \etal (2006), Wechsler \etal (2006), Gao \&
White (2007). This caveat notwithstanding, analytic models can provide
an elegant route to  capturing the  essential features of the
environmental dependence.

In standard galaxy formation models, the correlation between the
haloes and their large scale environment introduces a correlation
between the properties of galaxies and the regions they occupy
(e.g. Navarro, Abadi \& Steinmetz 2004; Abbas \& Sheth 2005, 2006;
Berlind \etal 2005). The observational   results of Skibba \etal
(2006), Blanton \etal (2006) and Tinker \etal (2007) support this
prediction and leave little room for a galaxy  assembly bias (see,
however, Croton \etal 2007).  Nevertheless, the influence of the
environmental dependence of halo properties on the galaxy population
remains unclear. It would be valuable to assess whether environmental
effects produced by the anisotropic collapse of haloes can leave a
detectable signature in  the properties of field galaxies.

\section*{Acknowledgements}

I wish to thank the anonymous referee for insightful comments; 
Avishai Dekel, Andr\'e Henriques, Yehuda Hoffman, Eyal
Neistein and Adi Nusser for helpful discussions; Havard Sandvik and
Geraint Harker for correspondence; and Noam  Libeskind for his
comments on a early version of this manuscript.  This work has been
supported by the German-Israel Einstein Centre and a Golda Meir
Fellowship at the Hebrew University.

\appendix

\section{Two-point correlations of the shear tensor}
\label{app:gcase}

\begin{figure}
\centering \resizebox{0.45\textwidth}{!}{\includegraphics{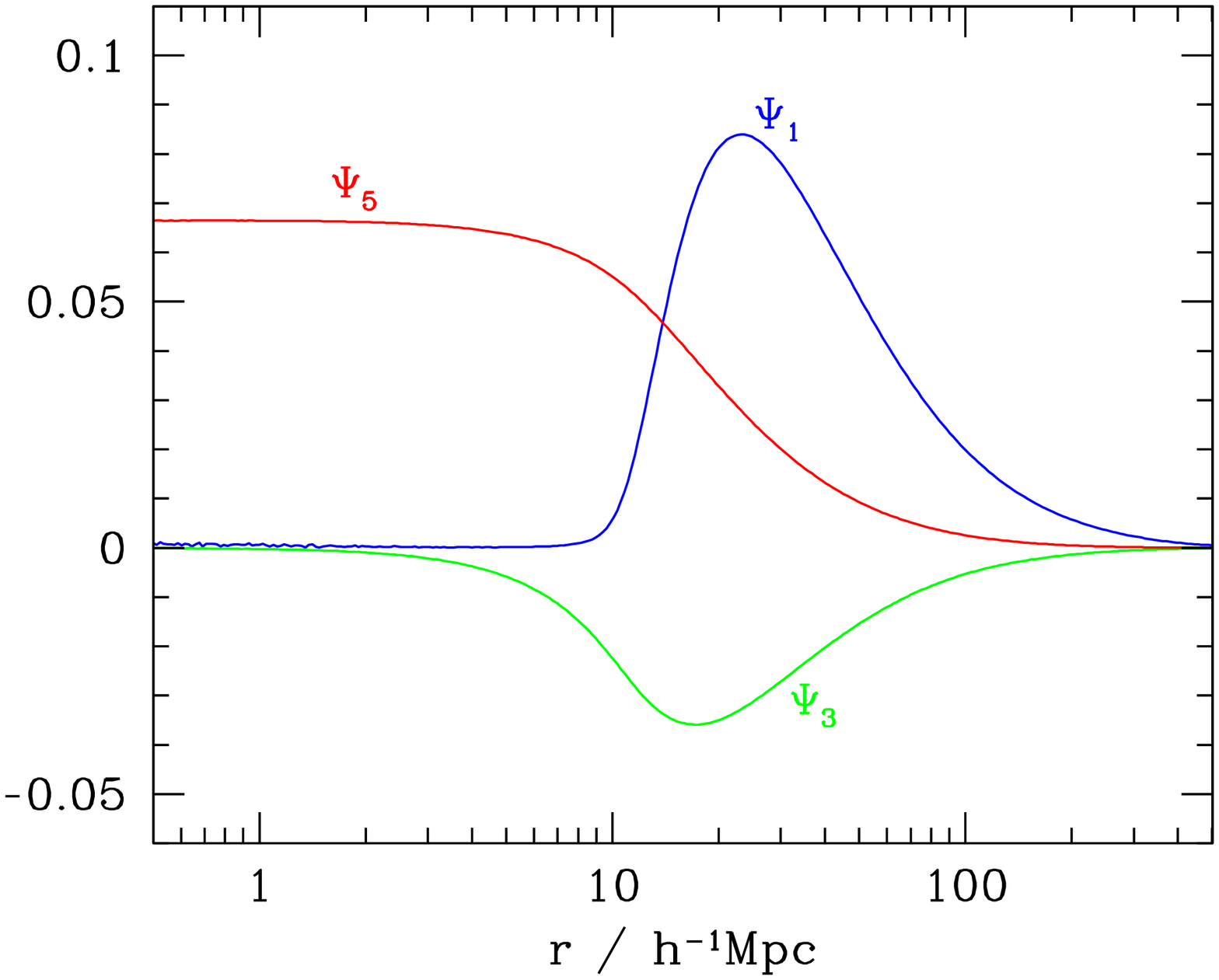}}
\caption{The correlation functions $\Psi_1(r)$, $\Psi_3(r)$ and
$\Psi_5(r)$ as a function of comoving separation $r$ for the
$\Lambda$CDM cosmology considered in this paper. They are normalised
by $\xi_{01}$ (eq.~\ref{eq:xi01}). The shear tensor at position $\vx$
and $\vx+\vr$ is smoothed on scale $R_0=10$ and $R_1=2\hmpc$,
respectively. Note that $\Psi_1$ is strongly suppressed on scale
$r\lsim R_0$.}
\label{fig:corrs}
\end{figure}

We consider the two-point correlation functions of an arbitrary
symmetric tensor field ${\rm T}_{ij}(\vx)$. Statistical isotropy and
symmetry imply  that, in position space, these correlations must be 
of the form
\begin{eqnarray}
\lefteqn{\la {\rm T}_{ij}(\vx){\rm T}_{lm}(\vx+\vr)\ra =
\Psi_1(r)\,\rh_i \rh_j \rh_l \rh_m} \nonumber \\ &&
+\,\Psi_2(r)\left(\rh_i \rh_l\delta_{jm}+\rh_i \rh_m\delta_{jl} +\rh_j
\rh_l\delta_{im}+\rh_j \rh_m\delta_{il}\right) \nonumber \\ &&
+\,\Psi_3(r)\left(\rh_i \rh_j\delta_{lm}+\rh_l \rh_m\delta_{ij}
\right)+\Psi_4(r)\,\delta_{ij}\delta_{lm} \nonumber \\ &&
+\,\Psi_5(r)\left(\delta_{il}\delta_{jm}+\delta_{im}\delta_{jl}
\right) \;,
\label{eq:corr1}
\end{eqnarray}
where $\rh_i=r_i/|\vr|$ and the functions $\Psi_i(r)$ depend on
$r=|\vr|$ only.  This is the most general ansatz for a symmetric,
isotropic correlation  tensor. In the case of a scalar (spin-0) tensor
such as the shear  $\xi_{ij}$ defined in eq.~(\ref{eq:strain}),
$\Psi_2=\Psi_3$ and  $\Psi_4=\Psi_5$. Explicit expression for the
functions $\Psi_i(r)$ can  be obtained from the Fourier transform of
$\la\xi_{ij}(\vx)\xi_{lm}(\vx+\vr)\ra$,
\begin{equation}
\la\xi_{ij}(\vk)\xi_{lm}(\vk)\ra = P_\delta(k)\,\kh_i \kh_j \kh_l
\kh_m\;,
\label{eq:corr2}
\end{equation} 
where $\kh_i=k_i/k$ and $P_\delta(k)$ is the power spectrum of the
density field $\delta(\vx)$. The following integrals are useful to the
calculation of $\Psi_i(r)$,
\begin{eqnarray}
\frac{1}{2}\int_{-1}^{+1}\!\!\dd\mu\,\mu^2\,e^{ikr\mu} \!\!\!\! &=&
\!\!\!\! \frac{1}{3}j_0(kr)-\frac{2}{3}j_2(kr) \nonumber \\
\frac{1}{2}\int_{-1}^{+1}\!\!\dd\mu\,\mu^4\,e^{ikr\mu} \!\!\!\! &=&
\!\!\! \frac{1}{5}j_0(kr)-\frac{4}{7}j_2(kr)+\frac{8}{35}j_4(kr)\;,
\label{eq:corr3}
\end{eqnarray}
where $j_\ell(x)$ are spherical Bessel functions of the first kind.
With these informations, the functions $\Psi_i$ may be conveniently
expressed as
\begin{eqnarray}
\Psi_1(r)\!\!\!\! &=& \!\!\!\!  \int_0^\infty\!\!\dd\ln
k\,\dkk\,j_4(kr) \\ \Psi_3(r)\!\!\!\! &=& \!\!\!\!
\int_0^\infty\!\!\dd\ln k\,\dkk\,\left[-\frac{1}{7}j_2(kr)-
\frac{1}{7}j_4(kr)\right] \nonumber \\ \Psi_5(r)\!\!\!\! &=& \!\!\!\!
\int_0^\infty\!\!\dd\ln k\,\dkk\,\left[\frac{1}{15}j_0(kr)+
\frac{2}{21}j_2(kr)+\frac{1}{35}j_4(kr)\right] \nonumber
\label{eq:psif}
\end{eqnarray}
In the limit $r\rightarrow 0$, both $\Psi_1$ and $\Psi_3$ vanish, but
$\Psi_5=\xi_{01}/15$ (Figure~\ref{fig:corrs}).  Notice that the
functions $\Psi_i$ can also be cast into the form given in,  e.g., Lee
\& Pen (2001), Crittenden \etal (2001), and Catelan \& Porciani (2001), 
which involves $J_n\equiv n r^{-n}\int_0^r\!\!\dd s\,\xi(s)s^{n-1}$ 
(this is best seen  by integrating $J_n$ by part).

\section{An integral over the SO(3) manifold}
\label{app:so3}

In this Appendix, we present the calculation of the integral
eq.~(\ref{eq:iso3}) over the special orthogonal group SO(3), where
$\alpha$ and $\lambda$ are real symmetric matrices.

Let $\vrr\equiv\vrr_{_{\hat{n}}}\!(\varphi)$ be the rotation by an
angle $\varphi$  about the axis $\nvh$. Since the Haar measure
$\dd\vrr$ is invariant under  both left and right transformations, we
can redefine $\vrr$ so as to  diagonalise $\lambda$ and $\alpha$. We
will therefore assume that the  matrices $\lambda$ and $\alpha$ are
diagonal, $\lambda={\rm diag}(\lambda_i)$ and $\alpha={\rm
diag}(\alpha_i)$.  It is worth noting that $\alpha$ and $\lambda$ are
not orthogonal matrices, but belong  to the (real) linear group
GL(3,R). Consider now the rotation of $\pi$  around the first axis,
$\sigma_1={\rm diag}(1,-1,-1)$, and its distinct  permutations
$\sigma_2={\rm diag}(-1,1,-1)$ and  $\sigma_3={\rm
diag}(-1,-1,1)$. Since the compositions $\sigma_i\vrr$ characterise
equivalent principal axis frames, the integral (\ref{eq:iso3})  should
in principle be performed over the quotient space
SO(3)/{$\{\sigma_1,\sigma_2,\sigma_3\}$}.  However, the $\sigma_i$s
are diagonal and we find
\begin{equation}
\tr\left[\left(\sigma_i\vrr\right)\lambda\left(\sigma_i\vrr\right)^\top
\!\alpha\right]=\tr\left(\vrr\lambda\vrr^\top\!\alpha\right)\;.
\label{eq:rotframe}
\end{equation}
Therefore, one can also perform the integration over the whole
orthogonal  group SO(3) and multiply the final result by $1/4$.

Integrals of the form
\begin{equation}
F(\alpha,\lambda)=\int_{{\rm G}}\!\!\dd\Omega\,
e^{\beta\tr\left(\Omega\lambda\Omega^{-1}\!\alpha \right)}\,
\label{eq:igk}
\end{equation}
where G ($\Omega\in$G) is a compact Lie group and $\alpha,\lambda$
belong to the linear group GL, appear in  Quantum Field Theory,
chemistry as well as in harmonic analysis (e.g.  Hua 1963; Itzykson \&
Zuber 1980; Wei \& Eichinger 1989). When G is the  unitary group U(N),
group theoretical techniques such as character  expansion can be used
to obtain a closed form determinantal evaluation of (\ref{eq:igk})
(e.g. Harish Chandra 1958; Balantekin 2000). Unfortunately, these
methods cannot be easily applied when G is the orthogonal group SO(N),
essentially because the representations of SO(N) and GL(N,R) are very
different. Instead, one generally relies on a specific parametrisation
of the rotation matrices and express the result in terms of a series
of orthogonal functions.

\begin{table*}
\caption{Quadrupole Wigner D-functions  ${\cal
D}^2_{_{m_1,m_2}}\!\left(\varphi,\vartheta,\psi\right)$ (in ZXZ
representation. Harmonics with $m_1,m_2=\pm 1$ are not shown).}
\begin{center}
\begin{tabular}{llll} \hline\hline
&  $m_2=-2$ & $m_2=0$ & $m_2=2$ \\ \hline $m_1=-2$ &
$\frac{1}{4}\left(1+\cos\vartheta\right)^2 e^{2i\varphi+2i\psi}$ &
-$\sqrt{\frac{3}{8}}\sin^2\!\vartheta\,  e^{2i\varphi}$ &
$\frac{1}{4}\left(1-\cos\vartheta\right)^2 e^{2i\varphi-2i\psi}$ \\
$m_1=0$ & -$\sqrt{\frac{3}{8}}\sin^2\!\vartheta\,  e^{2i\psi}$ &
$\frac{1}{2}\left(3\cos^2\!\vartheta-1\right)$ &
-$\sqrt{\frac{3}{8}}\sin^2\!\vartheta\,e^{-2i\psi}$ \\ $m_1=2$ &
$\frac{1}{4}\left(1-\cos\vartheta\right)^2 e^{-2i\varphi+2i\psi}$  &
-$\sqrt{\frac{3}{8}}\sin^2\!\vartheta\, e^{-2i\varphi}$ & $\frac{1}{4}
\left(1+\cos\vartheta\right)^2 e^{-2i\varphi-2i\psi}$ \\ \hline\hline
\end{tabular}
\end{center}
\label{table:table1}
\end{table*}

Here we follow the approach outlined in Wei \& Ei\-chin\-ger (1990)
and  parametrise the special rotation matrices $\vrr$ in terms of the
Euler  angles $0\leq \varphi,\psi\leq 2\pi$, $0\leq\vartheta\leq\pi$,
\begin{equation}
\vrr= \left(\begin{array}{ccc} c_\psi c_\varphi-c_\vartheta s_\varphi
s_\psi &  c_\psi s_\varphi+ c_\vartheta c_\varphi s_\psi &  s_\psi
s_\vartheta \\ -s_\psi c_\varphi-c_\vartheta s_\varphi c_\psi &
-s_\psi s_\varphi+c_\vartheta c_\varphi c_\psi &  c_\psi s_\vartheta
\\ s_\vartheta s_\varphi & -s_\vartheta c_\varphi &  c_\vartheta \\
\end{array}\right) \;,
\label{eq:euler}
\end{equation}
where $c_\psi=\cos\psi$, $s_\varphi=\sin\varphi$ etc. Notice that this
representation becomes singular when $\vartheta=0$ or $\pi$ (such
singularities are expected owing to the topology of SO(3)). The
original integral~(\ref{eq:iso3}) over the SO(3) manifold can be
written as a triple integral with the (normalised) invariant measure
$\dd\vrr=1/(8\pi^2)\sin\vartheta\,\dd\varphi\dd\vartheta\dd\psi$,
\begin{equation}
F(\alpha,\lambda)=\frac{1}{8\pi^2} \int_0^{2\pi}\!\!\dd\varphi
\int_0^\pi\!\!\dd\vartheta\,\sin\vartheta \int_0^{2\pi}\!\!\dd\psi\,
e^{\beta\tr\left(\vrr\lambda\vrr^\top\!\!  \alpha\right)}\;.
\label{eq:form0}
\end{equation}
Since the integrand is invariant under the transformations
$(\alpha,\lambda)\rightarrow \left(k\,\alpha,\frac{1}{k}\,\lambda
\right)$ and $(\alpha,\lambda)\rightarrow \left(\alpha+k\,\vii,
\lambda-\frac{k\,\tr\lambda}{k+\tr\alpha}\,\vii\right)$, where $k$ is
a real number and $\vii$ is the $3\times 3$ identity matrix,
$F(\alpha,\lambda)$ can be expressed as a function of four variables
instead of the original six matrix eigenvalues. This can also seen by
rewriting the trace as (Wei \& Eichinger 1990)
\begin{equation}
\tr\left(\vrr\lambda\vrr^\top\!\!\alpha\right)=
\frac{1}{3}\left(\tr\alpha\,\tr\lambda-\tr\tilde{\alpha}\,\tr\tilde{\lambda}
\right)+\tr\left(\vrr\tilde{\lambda}\vrr^\top\!\tilde{\alpha}\right)\;.
\label{eq:trace0}
\end{equation}
where, for instance, $\tilde{\alpha}={\rm diag}\left(\alpha_{13},
\alpha_{23},0\right)$, $\tilde{\lambda}={\rm diag}\left(\lambda_{13},
\lambda_{23},0\right)$, $\alpha_{ij}=\alpha_i-\alpha_j$ and
$\lambda_{ij}=\lambda_i-\lambda_j$. We enforce the ordering $\alpha_1
\geq\alpha_2\geq\alpha_3$ and $\lambda_1\geq\lambda_2\geq\lambda_3$,
so  that $\alpha_{13},\alpha_{23}\geq 0$ and
$\lambda_{13},\lambda_{23}\geq 0$.  The integral depends now on four
distinct combinations of the $\alpha$s and $\lambda$s:
$\epsi_+=(1/3)\,\tr\alpha\,\tr\lambda$, $\epsi_-=(1/3)\,
\tr\tilde{\alpha}\,\tr\tilde{\lambda}$, $\epsi_\alpha=\alpha_{12}/
\tr\tilde{\alpha}$, $\epsi_\lambda=\lambda_{12}/\tr\tilde{\lambda}$.
With this parametrisation, $-\infty\leq\epsi_+\leq\infty$,
$\epsi_-\geq 0$  and $0\leq\epsi_\alpha,\epsi_\lambda\leq 1$.

We can expand the trace  in terms of the Wigner D-functions ${\cal
D}^l_{_{m_1,m_2}}$, $l$ being the index of the representation (see,
e.g., Sakurai 1985). These 3D  harmonics generate irreducible
representations of the three-dimensional rotation group and,
therefore, form  a complete orthogonal set of functions defined on
SO(3) itself. Unsurprisingly, only  the quadrupole ($l=2$) rotation
matrices appear in the trace decomposition,
\begin{eqnarray}
\lefteqn{\tr\left(\vrr\tilde{\lambda}\vrr^\top\!\tilde{\alpha}\right)=
\frac{\epsi_-}{2}\left[\frac{2}{3}+{\cal D}^2_{_{0,0}}\right.}
\nonumber  \\ && - \left.\sqrt{\frac{3}{2}}\epsi_\alpha\left({\cal
D}^2_{_{0,-2}} +{\cal
D}^2_{_{0,2}}\right)-\sqrt{\frac{3}{2}}\epsi_\lambda \left({\cal
D}^2_{_{-2,0}}+{\cal D}^2_{_{2,0}}\right)\right.\nonumber \\  && +
\left.\frac{3}{2}\epsi_\alpha\epsi_\lambda\left({\cal D}^2_{_{-2,-2}}
+{\cal D}^2_{_{2,2}}+{\cal D}^2_{_{-2,2}}+{\cal D}^2_{_{2,-2}} \right)
\right]\;.
\label{eq:trace1}
\end{eqnarray}
The explicit form of these $l=2$ harmonics is given in
Table~\ref{table:table1}. Note that
$\tr(\vrr\tilde{\lambda}\vrr^\top\!\tilde{\alpha})$ depends on the
three ``shape'' parameters $\epsi_-$, $\epsi_\alpha$ and
$\epsi_\lambda$ solely, because points on SO(3) truly have only  three
degrees of freedom. The integral over the variable  $\psi$ can be
performed using the following identity (see~\S 3.937 of  Gradshteyn \&
Ryzbik 2000),
\begin{equation}
\frac{1}{2\pi}\int_0^{2\pi}\!\!\dd\psi\,e^{\beta\left[x\cos  (2\psi)+y
\sin (2\psi)\right]}=I_0\left(\sqrt{x^2+y^2}\right) \;,
\end{equation}
where $I_0$ is a modified Bessel function of the first kind. The
result  can be written
\begin{equation}
F(\alpha,\lambda)=e^{\beta\epsi_+}\,
W\left(\beta\epsi_-,\epsi_\alpha,\epsi_\lambda\right)\;.
\label{eq:form1}
\end{equation}
where the function
$W\left(\beta\epsi_-,\epsi_\alpha,\epsi_\lambda\right)$  is a double
integral which can be arranged such that
\begin{eqnarray}
\lefteqn{W\left(\beta\epsi_-,\epsi_\alpha,\epsi_\lambda\right)}
\nonumber \\  &=& \!\!\! e^{-\beta\epsi_-}\,\left\{\frac{1}{2\pi}
\int_0^1\!\!\dd r\int_0^{2\pi}\!\!\dd\varphi\,
\exp\left[\frac{3\beta\,\epsi_-}{4}\,g(r,\varphi,\epsi_\alpha) \right]
\nonumber \right. \\ && \left. \times I_0\left[\frac{3\beta\,
\epsi_-\epsi_\lambda}{4}\sqrt{h(r,\varphi,\epsi_\alpha)}\right]\right\}
\;,
\label{eq:wfunc}
\end{eqnarray}
where $r=-\cos\vartheta$. The functions $g(r,\varphi,\epsi_\alpha)$
and  $h(r,\varphi,\epsi_\alpha)$ are defined as
\begin{eqnarray}
g(r,\varphi,\epsi_\alpha)\!\!\!\! &=& \!\!\!\!
1+r^2+\epsi_\alpha\left(1-r^2\right)\cos(2\varphi)\nonumber \\
h(r,\varphi,\epsi_\alpha)\!\!\!\! &=& \!\!\!\!  g^2
-4\left(1-\epsi_\alpha^2\right)r^2 \;.
\end{eqnarray}
They are periodic of period $\pi$ in the argument $\varphi$.
Furthermore, on the domain defined by $0\leq r\leq 1$ and
$0\leq\varphi \leq 2\pi$, the function $g$ is bounded by
$1-\epsi_\alpha\leq g(r,\varphi, \epsi_\alpha)\leq
1+\epsi_\alpha$. Consequently, the double integral that appears in
eq.~(\ref{eq:wfunc}) is always larger than or equal to 1. The equality
holds only when $\alpha$ and/or $\lambda$ is proportional to the
identity matrix.

Expanding the integrand of (\ref{eq:wfunc}) about $\beta\epsi_-=0$
gives,  upon integration, the following series
\begin{eqnarray}
\lefteqn{W\left(\beta\epsi_-,\epsi_\alpha,\epsi_\lambda\right)} \\ &&
\approx 1+\frac{\beta^2\epsi_-^2}
{40}p_1(\epsi_\alpha)p_1(\epsi_\lambda)+\frac{\beta^3\epsi_-^3}
{840}p_2(\epsi_\alpha)p_2(\epsi_\lambda)\nonumber\\ &&
+\frac{\beta^4\epsi_-^4}{4480}p_1^2(\epsi_\alpha)p_1^2(\epsi_\lambda)
+\frac{\beta^5\epsi_-^5}{73920}p_1(\epsi_\alpha)p_2(\epsi_\alpha)
p_1(\epsi_\lambda)p_2(\epsi_\lambda)\nonumber
\label{eq:iwf}
\end{eqnarray}
where $p_1(x)=(3x^2+1)$ and $p_2(x)=(9x^2-1)$.  We have found that
this truncated, fifth-order expansion is accurate to  within 2 per
cent in the range $0\leq\beta\epsi_-\lsim 1.5$ and can be used
efficiently in the computation of the group integral~(\ref{eq:iso3}).
Conditional probability distributions for the relative orientation can
be derived in a straightforward way from the preceding results.

Noteworthy is the special case G=SO(2), in which the integral
(\ref{eq:igk}) can be written in closed form,
\begin{equation}
F(\alpha,\lambda)=e^{\beta\epsilon_+}
I_0\left(\frac{1}{2}\alpha_{12}\lambda_{12}\right)\;,
\label{eq:iso2}
\end{equation}
where the parameter $\epsi_+$ is now
$\epsi_+=(1/2)\,\tr\alpha\,\tr\lambda$.

\section{Potential of a homogeneous ellipsoid}
\label{app:pot}

For a homogeneous ellipsoid defined such that $\rho(\vr)=\delta\rho$
if  $\vr^\top\left(\vaa\vaa^\top)^{-1}\right)\vr\leq 1$, and zero
otherwise,  the potential at any exterior point
$\vr=\left(r_1,r_2,r_3\right)$ is  (Kellog 1929; Chandrasekhar 1969)
\begin{equation}
\Phi_e(\vr)=\pi G\delta\rho\,\left[C(\rho)-\sum_i
b_i(\rho)\,r_i^2\right]\;,
\label{eq:potext}
\end{equation}
while the potential inside and on the boundary of the ellipsoid is
\begin{equation}
\Phi_i(\vr)=\pi G\delta\rho\,\left[C(0)-\sum_i b_i(0)\,r_i^2\right]\;.
\label{eq:potint}
\end{equation}
$AA^\top$ is a positive definite matrix whose eigenvalues are the
square  of the principal axis lengths $A_1\geq A_2\geq A_3>0$. The
variable $\rho$ is defined as the algebraically largest root of the
following cubic equation,
\begin{equation}
\frac{x^2}{A_1^2+\rho}+\frac{y^2}{A_2^2+\rho}+\frac{z^2}{A_3^2+\rho}=1\;.
\label{eq:cubic}
\end{equation}
The functions $C(\rho)$ and $b_i(\rho)$ are given by
\begin{eqnarray}
C(\rho) \!\!\!\! &=& \!\!\!\! A_1 A_2 A_3\,\int_{\rho}^\infty\!
\frac{\dd u}{\Delta(u)} \nonumber \\ b_i(\rho) \!\!\!\! &=& \!\!\!\!
A_1 A_2 A_3\,\int_{\rho}^\infty\!  \frac{\dd
u}{\left(A_i^2+u\right)\Delta(u)}\;,
\label{eq:bell}
\end{eqnarray}
where $\Delta(u)=\Pi_{k=1}^3\left(A_k^2+u\right)^{1/2}$. These
functions  can be expressed in terms of the Legendre's incomplete
elliptic integrals  of first and second kind (Binney \& Tremaine
1987). The gravitational  potential energy of this self-gravitating
ellipsoid is
\begin{equation}
E_{\rm pot}=-\frac{3}{10}\,GM^2  \int_0^\infty\!\!\frac{\dd
u}{\Delta(u)}
\label{eq:pot}
\end{equation}
and is proportional to the potential at $\vx=0$, $C(0)$. Notice also
that Poisson's equation implies $\sum_i b_i(0)=2$. Eq.~(\ref{eq:pot})
corrects a typographical error in eq.~(24) of Eisenstein \& Loeb 
(1995).

\label{lastpage}

\end{document}